\documentclass[preprint]{aastex}

\shorttitle{Numerical Method for GRMHD}
\shortauthors{De Villiers and Hawley}

\begin{document}

\title{A Numerical Method for General Relativistic Magnetohydrodynamics}


\author{Jean-Pierre De Villiers and John F. Hawley}
\affil{Astronomy Department\\
University of Virginia\\ 
P.O. Box 3818, University Station\\
Charlottesville, VA 22903-0818}
\email{jd5v@virginia.edu; jh8h@virginia.edu}

\begin{abstract}
This paper describes the development and testing of a general relativistic
magnetohydrodynamic (GRMHD) code to study ideal MHD in the fixed background of a
Kerr black hole.  The code is a direct extension of the hydrodynamic code of
Hawley, Smarr, and Wilson, and uses Evans and Hawley constrained transport (CT) to
evolve the magnetic fields.  Two categories of test cases were undertaken.  A
one dimensional version of the code (Minkowski metric) was used to verify code
performance in the special relativistic limit.  The tests include Alfv\'en wave
propagation, fast and slow magnetosonic shocks, rarefaction waves, and both
relativistic and non-relativistic shock tubes.  A series of one- and
two-dimensional tests were also carried out in the Kerr metric:  magnetized
Bondi inflow, a magnetized inflow test due to Gammie, and two-dimensional
magnetized constant-$l$ tori that are subject to the magnetorotational
instability.
\end{abstract}



\section{Introduction}

Accretion into black holes is believed to account for a wide variety of
astrophysical phenomena, from solar-mass black holes in X-ray binaries, to
supermassive black holes in active galactic nuclei.  The basic theory of black
hole accretion was laid out more than thirty years ago in a number of
now-classic papers (e.g.  Novikov \& Thorne 1972; Shakura \& Sunyaev 1973;
Lynden-Bell and Pringle 1974).  Since then, theoretical progress has been
steady, but perhaps at a slower rate than might have been hoped.  The problem is
complex, involving time-dependent, three-dimensional dynamics of magnetized
plasmas in the relativistic potential of Kerr black holes.  Solutions are
impossible to obtain analytically beyond simplified cases that rely upon
time-stationarity and spatial symmetry.  Because of this, numerical experiments
must play an increasingly important role in driving theory.  The importance of
numerical simulations in the investigation of such complex physics has long been
recognized, but until recently the necessary computational hardware was not
available.

We are now in a position to pursue full 3D simulations with general relativistic
MHD codes.  Again, the basics were laid out thirty years ago by Wilson (1972)
who pioneered both hydrodynamic and magnetohydrodynamic (MHD) two-dimensional
simulations of black hole accretion (Wilson 1975; 1977; 1978).  Since then there
have been a significant number of efforts by a number of groups to develop
accretion simulations for both Newtonian and relativistic gravitational fields,
and for both hydrodynamics and magnetohydrodynamics.  This paper details our
current effort to develop a fully general relativistic MHD accretion code.

Why focus on general relativistic MHD?  In fact, for many applications Newtonian
gravity (or a pseudo-Newtonian mock-up of a black hole) is sufficient.  In some
important cases, however, it is not.  The detailed physics of gas interacting
with a black hole provides, in principle, a rare test of strong-field general
relativity, but only if we understand the unique properties of accretion in the
fully relativistic case.  Distinctive and astrophysically interesting effects
are expected from the Kerr metric, including those due to the lack of a stellar
surface, the presence of an innermost stable circular orbit, and the dragging of
inertial frames.  The need to include MHD in an accretion simulation is also now
quite clear.  Magnetic fields play an essential role in the outward transport of
angular momentum in accretion disks through the action of the magnetorotational
instability (MRI; Balbus \& Hawley 1998).  This, in turn, means the simulations
must be three-dimensional, for only in three dimensions can there be a
self-sustaining (and accretion-sustaining) MHD dynamo.

In this paper we describe an algorithmic approach to the three-dimensional
equations of general relativistic MHD that is based upon techniques first
developed for axisymmetric hydrodynamics around black holes (Hawley, Smarr \&
Wilson 1984a, 1984b; hereafter HSWa and HSWb).  The hydrodynamic portion of the
code was recently tested and employed in three-dimensional simulations in the
Kerr metric by De Villiers \& Hawley (2002).  Here we cast the equations of
relativistic MHD in a form that is compatible with the HSW formulation.  We will
describe how those equations are evolved numerically.  We will also discuss a
suite of tests suitable for relativistic codes, and show how our code performs
against those tests.

It is important to recognize that the development of numerical algorithms in any
area of physics with few analytic solutions, and even fewer experimental
observations, is likely to be difficult.  Progress will be made only by
exploring a variety of approaches.  Such a situation necessarily precludes the
development of only one numerical algorithm, or even a single code
implementation of a promising algorithm.  Some approaches will have strengths
for certain applications, but it is unlikely that one numerical technique will
be optimal for all problems.  In any case, experience gained, both with what
works and what does not, will be essential as we work toward developing a
consensus within the community.  With this in mind, our own efforts have
concentrated on extending the established general relativistic hydrodynamic
approach of HSW, since our main area of interest, that of accretion flows, has
been well handled (in the hydrodynamic case) by this solver, and by similar
algorithms designed to perform Newtonian MHD simulations.  Our test suite has
been developed in conjunction with Gammie, McKinney, \& T\'oth (2002), who use
an alternative scheme for general relativistic MHD.

The plan of this paper is as follows.  In \S2 we write the equations of general
relativistic MHD in the form that they will be solved in the code.  In \S3 we
describe the numerical implementation of our algorithm.  In \S4 we describe the
one-dimensional Minkowski (special relativistic) test problems.  In \S5 we
describe the test problems in the Kerr and Schwarzschild metrics, and present
our test results.  The summary and conclusions are presented in \S6.

\section{Equations of General Relativistic Magneto-Hydrodynamics}

We wish to study the evolution of a magnetized fluid in the background
spacetime metric of a Kerr black hole.  We work in
Boyer-Lindquist coordinates, for which the line element has the form,
\begin{equation}\label{kerr}
{ds}^2=g_{t t}\,{dt}^2+2\,g_{t \phi}\,{dt}\,{d \phi}+g_{r r}\,{dr}^2+
 g_{\theta \theta}\,{d \theta}^2+g_{\phi \phi}\,{d \phi}^2.
\end{equation}
In keeping with Misner, Thorne, \& Wheeler (1973), we use the metric
signature $(-,+,+,+)$, along with geometrodynamic units where $G = c =
1$; the black hole mass is unity, $M=1$.  The determinant of the
4-metric is $g$, and $\sqrt{-g} = \alpha\,\sqrt{\gamma}$ where $\alpha$
is the lapse function, $\alpha=1/\sqrt{-g^{tt}}$, and $\gamma$ is the
determinant of the spatial $3$-metric.

The state of the relativistic test fluid at each point in the spacetime
is described by its density, $\rho$, internal energy, $\epsilon$,
$4$-velocity $U^\mu$, and isotropic pressure, $P$, which is related to
the first two scalars via the equation of state of an ideal gas,
$P=\rho\,\epsilon\,(\gamma-1)$, where $\gamma$ is the adiabatic
exponent.

The equations of general relativistic MHD are the law of 
baryon conservation,
\begin{equation}\label{barcons}
{\nabla}_{\mu}\,(\rho\,U^{\mu}) = 0 ,
\end{equation}
where ${\nabla}_{\mu}$ is the covariant derivative, 
the conservation of stress-energy,
\begin{equation}\label{tmncons}
{\nabla}_{\mu}{T}^{\mu\,\nu} = 0 ,
\end{equation}
where ${T}^{\mu\,\nu}$ is the energy-momentum tensor for the fluid, 
and Maxwell's equations,
\begin{eqnarray}
\label{maxwell.1}
{\nabla}_{\mu}{F}^{\mu\,\nu}  & = & 4\,\pi\,J^\nu ,\\
\label{maxwell.2}
{\nabla}_{\mu}{}^*{F}^{\mu\,\nu}  & = & 0,
\end{eqnarray}
where $F^{\mu \nu}$ is the electromagnetic field strength tensor, 
$J^\mu = (\rho_c,J^i)$ is the current $4$-vector, and 
$\rho_c$ the charge density. 
The dual tensor is defined as
\begin{eqnarray}
\label{dual.2}
{}^*F^{\mu\,\nu} & = & {1 \over 2}\,\epsilon^{\mu\,\nu\,\delta\,\gamma} 
 F_{\delta \gamma} ,
\end{eqnarray}
where $\epsilon^{\mu\,\nu\,\delta\,\gamma}
= -(1/\sqrt{-g})\,[\mu\,\nu\,\delta\,\gamma]$ is the contra-variant
form of the Levi-Civita tensor. Maxwell's equations are supplemented 
by the equation of charge conservation ${\nabla}_{\mu}J^\mu = 0$.

The energy-momentum tensor consists of perfect fluid and 
electromagnetic parts, ${T}^{\mu\,\nu} \equiv {T}^{\mu\,\nu}_{(fluid)}+
{T}^{\mu\,\nu}_{(EM)}$, and has the form
\begin{equation} \label{tmndef}
{T}^{\mu\,\nu} = \rho\,h\,{U}^{\mu}\,{U}^{\nu}+
 P\,{g}^{\mu\,\nu} + {1 \over 4\,\pi} \left({F^\mu}_\alpha\,F^{\nu \alpha} 
 - {1 \over 4} F_{\alpha \beta}\,F^{\alpha \beta}\,g^{\mu \nu} \right) .
\end{equation}
Following Lichnerowicz (1967) and Anile (1989), we define the magnetic 
induction and the electric field in the rest frame of the fluid,
\begin{eqnarray}
\label{bdef.2}
B^{\mu} & = & {}^*F^{\mu\,\nu}\,U_\nu ,\\
\label{edef.2}
E^{\mu} & = & F^{\mu \nu}\,U_\nu .
\end{eqnarray}
We adopt the ideal MHD limit and assume infinite conductivity (the
flux-freezing condition), wherein the electric field in the fluid rest
frame is zero, i.e., $F^{\mu \nu}\,U_\nu = 0$.

We combine the definition of the magnetic induction (\ref{bdef.2}) with 
(\ref{dual.2}) and the condition for infinite conductivity to obtain
\begin{equation}
\label{fmunudef.2}
F_{\mu\,\nu} = \epsilon_{\alpha\,\beta\,\mu\,\nu}\,B^{\alpha}\,U^\beta ,
\end{equation}
where 
$\epsilon_{\mu\,\nu\,\delta\,\gamma}=\sqrt{-g}\,[\mu\,\nu\,\delta\,\gamma]$.
The orthogonality 
condition 
\begin{equation}\label{bnorm}
B^{\mu}\,U_{\mu} = 0
\end{equation}
follows directly from (\ref{bdef.2}) and the anti-symmetry of $F^{\mu \nu}$.

Using these results, it is possible to rewrite the 
electro-magnetic portion of the energy-momentum tensor as
\begin{equation} 
\label{tmunuem}
{T}^{\mu\,\nu}_{(EM)} = \left(
{1 \over 2}\,g^{\mu \nu}\,{\|b\|}^2+U^\mu\,U^\nu\,{\|b\|}^2-
 b^\mu\,b^\nu \right) ,
\end{equation}
where $b^\mu \equiv B^\mu/\sqrt{4\,\pi}$, and 
${\|b\|}^2=g^{\mu\,\nu}\,b_\mu\,b_\nu$.

The induction equation (\ref{maxwell.2}) is solved using the 
equivalent form 
\begin{equation}\label{ctinduction}
\partial_\delta\,F_{\alpha \beta} + 
\partial_\alpha\,F_{\beta \delta} +
\partial_\beta\,F_{\delta \alpha } = 0 .
\end{equation}
Working directly with
$F_{\mu\nu}$ replaces covariant derivatives with simple coordinate
derivatives.  This is the basis for the Constrained Transport (CT) 
formulation of Evans \& Hawley (1988; hereafter EH88) for 
the induction equation.
In CT we  make the identification
\begin{equation}
{\cal{B}}^r      = F_{\phi \theta} \, , \,
{\cal{B}}^\theta = F_{r \phi} \, , \,
{\cal{B}}^\phi   = F_{\theta r} ,
\end{equation}
where ${\cal{B}}^j$ are the CT magnetic field variables.
With $t$ for one of the indices in (\ref{ctinduction}) we get the
evolution equations for ${\cal{B}}^j$, while
permuting over the spatial indices yields
$\partial_r\,F_{\phi \theta} + 
\partial_\theta\,F_{r \phi} +
\partial_\phi\,F_{\theta r } = 0, $
which is the familiar divergence-free condition.  
The CT magnetic field is considered the fundamental
expression of the magnetic field for evolution in the induction
equation.  Equation (\ref{fmunudef.2}) relates the CT field 
to the field $B^{\mu}$ (or $b^{\mu}$) which is
fundamental in the definition of ${T}^{\mu\,\nu}_{(EM)}$.

The induction equation (\ref{maxwell.2}) can also be 
rewritten by substituting definitions,
\begin{equation}\label{max2.b}
{\nabla}_{\alpha}\left(U^\alpha\,b^\beta- b^\alpha\,U^\beta\right) =0 .
\end{equation}
By expanding this equation using the product rule and applying the 
orthogonality condition (\ref{bnorm}), we obtain the identity
\begin{equation}\label{bident}
U_\nu\,b^\mu{\nabla}_{\mu}\,U^\nu =0 .
\end{equation}
Proofs of the results for (\ref{fmunudef.2}), (\ref{tmunuem}), and
(\ref{max2.b}) can be found in Appendix A. 

In preparation for discretizing these equations we list the
specific variables that will be used in the code.  First, the transport 
velocity (also known as the coordinate velocity) $V^\mu$ is defined
\begin{equation}\label{tranvel}
 V^\mu = {U^\mu \over U^t} ,
\end{equation}
where $U^t = W/\alpha$, and $W$ is the relativistic gamma-factor,
which can be expressed as
\begin{equation}\label{gred}
 W = \left[g^{tt}\left(g_{tt} + 2\,g_{t \phi}\,V^\phi +
 g_{r r}\,{(V^r)}^2 + g_{\theta \theta}\,{(V^\theta)}^2 + 
 g_{\phi \phi}\,{(V^\phi)}^2\right)\right]^{-1/2},
\end{equation}
a result which follows directly from the definition of transport 
velocity and the normalization condition on the $4$-velocity,
\begin{equation}\label{vnorm}
 g_{\mu \nu}\,U^\mu\,U^\nu = -1.
\end{equation}

The magnetic field of the fluid is described by two sets of variables,
the constrained transport (CT) magnetic field, ${\cal{B}}^i$, and
magnetic field $4$-vector $b^\mu$.  The latter is fundamental to the
definition of the total four momentum,
\begin{equation}\label{momdef}
 S_\mu = (\rho\,h\ + {\|b\|}^2)\,W\,U_\mu ,
\end{equation}
and the  normalization condition
\begin{equation}\label{momnorm}
g^{\mu \nu}\,S_\mu\,S_\nu = -{\left[(\rho\,h+{\|b\|}^2)\,W\right]}^2 ,
\end{equation}
which is algebraically equivalent to (\ref{vnorm}). 
Finally, we define auxiliary density and energy functions $D =
\rho\,W$ and $E = D\,\epsilon$.  The set of variables $D$, $E$,
$S_\mu$, ${\cal{B}}^i$, $V^i$, and $b_\mu$ will be those on which the
numerical scheme is built.

Since we are working in a coordinate basis, we also record here
the identities for the divergence of a four-vector and a tensor
\begin{equation}\label{divvec}
{\nabla}_{\mu}\left(f \,{v}^{\mu}\right) = {1 \over \sqrt{-g}}\,
\partial_{\mu}\left( \sqrt{-g} \,f \, {v}^{\mu}\right),
\end{equation}
\begin{equation}\label{divten}
{\nabla}_{\mu}\left({\beta}^{\mu\,\nu}\right) = {1 \over \sqrt{-g}}\,
\partial_{\mu}\left( \sqrt{-g}\,{\beta}^{\mu\,\nu}\right)+
{\Gamma}^{\nu}_{\epsilon\,\mu}\,{\beta}^{\mu\,\epsilon} .
\end{equation}
 
\subsection{The Equations of Baryon, Energy, and Momentum Conservation}

The equation of baryon conservation (\ref{barcons}) is unaltered by the 
presence of magnetic fields, and can be expanded in terms of the code 
variables to read (keeping in mind that the metric is stationary),
\begin{equation}\label{masscons}
\partial_t\,D + {1 \over
\sqrt{\gamma}}\,\partial_j\,(D\,\sqrt{\gamma}\,V^j)  = 0.
\end{equation}

The equation of energy conservation is derived by 
contracting (\ref{tmncons}) with ${U}_{\nu}$,
\begin{equation}\label{eq.1a}
{U}_{\nu}\,{\nabla}_{\mu}{T}^{\mu\,\nu} =
 {U}_{\nu}\,{\nabla}_{\mu}\left\{ 
 \left(\rho\,h+{\|b\|}^2\right)\,U^{\mu}\,U^{\nu}+
 \left(P+{{\|b\|}^2 \over 2}\right){g}^{\mu\,\nu}-
 b^\mu\,b^\nu\right\} = 0 .
\end{equation}
By using identity (\ref{bident}) and the law of baryon conservation
(\ref{barcons}), we recover the local energy conservation law 
\begin{equation}\label{econs}
{\nabla}_{\mu} \left(\rho\,\epsilon\,U^{\mu}\right)+
 P\,{\nabla}_{\mu} U^{\mu}= 0 ,
\end{equation}
which is unaffected by the presence of magnetic fields in the ideal
MHD limit and corresponds to an equation of entropy conservation.
Applying the definition for the auxiliary energy function $E$, 
the energy equation is rewritten as follows:
\begin{equation} \label{enfinal}
 \partial_{t}\left(E\right)+{1 \over \sqrt{\gamma}}\,
\partial_{i}\left(\sqrt{\gamma}\,E\,V^i\right)
+ P\,\partial_{t}\left(W\right) + 
{P\over\sqrt{\gamma}}\,\partial_{i}\left(\sqrt{\gamma}\,W\,V^i\right)
= 0 ,
\end{equation}
which is the same as in HSW.

The momentum conservation equations follow from
\begin{equation}\label{mom.1}
\nabla_\mu\,{T^\mu}_\nu = {\nabla}_{\mu}\left\{ 
 \left(\rho\,h+{\|b\|}^2\right)\,U^{\mu}\,U_{\nu}+
 \left(P+{{\|b\|}^2 \over 2}\right){\delta^\mu}_\nu-
 b^\mu\,b_\nu\right\} = 0 .
\end{equation}
Using the definition of momentum $S_\nu$ the first term in the
preceding expression can be rewritten as $S_\nu\,V^\mu/\alpha$ and  
simplified to ${S_\nu\,S^\mu / \alpha\,S^t}$,
and also using ${\Gamma_{\mu\,\nu}}^\gamma\,{\Pi_\gamma}^\mu =
 {1 \over 2}\,\Pi^{\gamma\,\mu}\,\partial_\nu\,g_{\mu\,\gamma} = 
-{1 \over 2}\,\Pi_{\gamma\,\mu}\,\partial_\nu\,g^{\mu\,\gamma}$, 
which holds for any symmetric tensor, we rewrite the momentum
equation as
\begin{equation}\label{mom.2}
{1 \over \alpha\,\sqrt{\gamma}}\,
\partial_\mu\,\sqrt{\gamma}\,S_\nu\,V^\mu +{1 \over 2\,\alpha}\,
{ S_\alpha\,S_\beta \over S^t }\,\partial_\nu\,g^{\alpha \beta}+
\partial_\nu\,\left(P+{{\|b\|}^2 \over 2}\right)-
{1 \over \alpha\,\sqrt{\gamma}}\,
\partial_\mu\,\alpha\,\sqrt{\gamma}\,b^\mu\,b_\nu-{1 \over 2}\,
 b_\alpha\,b_\beta\,\partial_\nu\,g^{\alpha \beta}
= 0 .
\end{equation}
To obtain the final form of the equations, multiply (\ref{mom.2})
by the lapse $\alpha$, split the $\mu$ index into
its space ($i$) and time ($t$) components, and restrict $\nu$ to the
spatial indices ($j$) only:
\begin{equation}\label{mom.3}
\partial_t\left(S_j-\alpha\,b_j\,b^t\right)+
  {1 \over \sqrt{\gamma}}\,
  \partial_i\,\sqrt{\gamma}\,\left(S_j\,V^i-\alpha\,b_j\,b^i\right)+
  {1 \over 2}\,\left({S_\epsilon\,S_\mu \over S^t}-
  \alpha\,b_\mu\,b_\epsilon\right)\,
  \partial_j\,g^{\mu\,\epsilon}+
  \alpha\,\partial_j\left(P+{{\|b\|}^2 \over 2}\right) = 0 .
\end{equation}
The $\nu$ index can be restricted to the spatial indices because the
equation that arises from $\nu=t$ for the time components of momentum
and magnetic fields is redundant, corresponding to a total energy
conservation equation.   In our formalism, we solve equation
(\ref{enfinal}) separately for the internal energy.  The time
component of $S_\mu$ is obtained from the normalization
condition on the momentum (\ref{momnorm}), and on the relations
between components of the magnetic field. 

We note that there are alternate ways to write this equation that,
while analytically equivalent, correspond to significant differences in
numerical implementation.  One is to evolve directly the total stress
energy $T^{\mu\nu}$ and solve algebraically for the primitive
variables.  This approach is taken by Koide, Shibata, \& Kudoh (1999),
Komissarov (1999), and Gammie, McKinney \& T\'oth (2002).  Yet another
alternative is to write the divergence of the electromagnetic
stress-energy tensor in the form $J_\mu F^{\mu j}$ (the familiar
$J\times B$ force from Newtonian MHD) and regard it as a source term
for the hydrodynamic momentum evolution; this approach was used by
Wilson (1978).

\subsection{The Induction Equation}

The equations of energy and momentum conservation have been expressed
in terms of the magnetic field $4$-vector $b^{\mu}$ since this yields
expressions structurally compatible with the existing hydrodynamic
equations.  This has important advantages in the numerical
implementation of these equations.  However, the induction equation is
evolved with the constrained transport (CT) approach of EH88 using the
variables ${\cal {B}}^i$. The practical consequence is that we must
translate between the two sets of variables.

Working with the second of Maxwell's equations, 
(\ref{maxwell.2}), we have previously identified the space-space components of
$F_{\mu\nu}$ with the CT field ${\cal{B}}^j$. It is easy to show that
we obtain the same result using (\ref{max2.b}),
\begin{equation}\label{ct.1}
{\nabla}_{\mu}\left(U^\nu\,b^\mu- b^\nu\,U^\mu\right) = 
\partial_\mu \sqrt{\gamma}\,W\,
 \left(V^\nu\,b^\mu- b^\nu\,V^\mu\right) = 0 .
\end{equation}
By splitting the above equation into two pieces and defining 
the CT magnetic field as 
\begin{equation}
\label{fmunudef.3}
{\cal{B}}^i=\sqrt{4\,\pi}\,\sqrt{\gamma}\,W\,(b^i-V^i\,b^t),
\end{equation}
it is possible to write
\begin{eqnarray}\label{ct.3a}
\partial_j \left({\cal{B}}^j\right) & = 0 & (\nu=0) ,\\
\label{ct.3b}
\partial_t \left({\cal{B}}^i\right) -
\partial_j \left(V^i\,{\cal{B}}^j-{\cal{B}}^i\,V^j\right)& = 0 & (\nu=i),
\end{eqnarray}
the starting point for the EH88 CT scheme. In addition,
the time-space components of $F_{\mu \nu}$ can be obtained by applying the
condition of infinite conductivity, $F^{\mu \nu}\,U_\nu = 0$, to obtain
\begin{equation}\label{maxwell.2c}
F_{t r}      = V^\phi\,{\cal{B}}^{\theta} 
-V^\theta\,{\cal{B}}^{\phi}  \, , \,
F_{t \theta} = V^r\,{\cal{B}}^{\phi} -V^\phi\,{\cal{B}}^{r} \, , \,
F_{t \phi}   = V^\theta\,{\cal{B}}^{r} - V^r\,{\cal{B}}^{\theta} .
\end{equation}
These can be identified with ${\cal{E}}^j$, the CT electromotive forces
(EMFs, EH88).

The divergence-free character of the magnetic field is maintained by
evolving the induction equation using these CT variables.  Hence 
the magnetic induction $b^\mu$ is derived from the CT magnetic field.
We obtain the
spatial components using (\ref{fmunudef.3}). To obtain $b^t$, we use
the definition of the magnetic induction (\ref{bdef.2}),
\begin{equation}\label{bdef.4}
b^{\mu} = {1 \over \sqrt{4 \pi}} {}^*F^{\mu\,\nu}\,U_\nu
 = {1 \over 2\,\sqrt{4 \pi}} \epsilon^{\mu\,\nu\,\delta\,\gamma}\,
 U_\nu\,F_{\delta\,\gamma} ,
\end{equation}
and we expand this definition and substitute for the Faraday tensor. 
This allows us to obtain the time-component of the magnetic field,
\begin{eqnarray}\label{bdef.5}
b^t & = & {W \over \sqrt{4 \pi} \alpha^2\,\sqrt{\gamma}}\, 
 \left({V^r\,{\cal{B}}^r \over g^{rr}}+
  {V^\theta\,{\cal{B}}^\theta \over g^{\theta \theta}}+
  {g^{tt}\,V^\phi - g^{t \phi} \over 
   g^{tt}\,g^{\phi \phi} - {(g^{t \phi})}^2}\,{\cal{B}}^\phi\right) .
\end{eqnarray}
This relationship can also be obtained  from the orthogonality condition
(\ref{bnorm}). Using this result and the relations in 
(\ref{fmunudef.3}) we obtain, after some algebra, a useful expression for 
$\|b^2\|$,
\begin{eqnarray}\label{bsqdef}
\|b^2\| & = & {1 \over 4\,\pi\,\gamma\, W^2}\, 
  \left({{{\cal{B}}^r}^2\over g_{rr}}+
        {{{\cal{B}}^\theta}^2 \over g_{\theta\theta}}+
        {g^{tt}\,{{\cal{B}}^\phi}^2 \over 
   g^{tt}\,g^{\phi \phi} - {(g^{t \phi})}^2}\right) \\
\nonumber
 & &  +{1 \over 4\,\pi\,\gamma\, \alpha^2}\,
 {\left({V^r\,{\cal{B}}^r \over g^{rr}}+
  {V^\theta\,{\cal{B}}^\theta \over g^{\theta \theta}}+
  {(g^{tt}\,V^\phi - g^{t \phi})\,{\cal{B}}^\phi\over 
   g^{tt}\,g^{\phi \phi} - {(g^{t \phi})}^2}\right)}^2
   .
\end{eqnarray}
The advantage of this form is that it guarantees that
$\|b^2\|$ is positive. The derivation is supplied in Appendix A.

\section{Implementing GRMHD}

The GRMHD code evolves time-explicit, operator-split, finite difference forms of
equations (\ref{masscons}), (\ref{enfinal}), (\ref{mom.3}), and (\ref{ct.3b}).
Variables are placed on a fixed spatial grid using Boyer-Lindquist coordinates.
They are advanced in time over a timestep size $\Delta t$ that remains fixed for
the duration of the simulation, and is determined by the extremal light-crossing
time for a grid zone, as described in HSWb.  The evolution algorithm is a
three-dimensional generalization of the solver described in HSWb extended to
include the contribution of magnetic fields.  The hydrodynamic portion of the
solver was described in De Villiers and Hawley (2002).

One timestep consists of three sequential sub-steps:  the induction
step, the transport step, and finally the source step.  The induction
step updates the magnetic field vector ${\cal{B}}^i$ using the
Constrained Transport framework of EH88.  This is
discussed in greater detail below.  After the update, the CT field is
transformed to the magnetic field in the fluid rest frame, $b^i$.
Velocities $V^i$ are computed using the normalization condition
during the source step.

Three versions of GRMHD have been used in development and testing:
(1) a 1D version with Minkowski metric to do Alfv\'en and shock wave
tests, described here in \S4, (2) a 2D axisymmetric version in the Kerr
metric to perform the tests described in \S5, and (3) the full 3D
production version of the code.  The 3D version of the code uses
message passing parallelism with domain decomposition, where the global
grid is partitioned into subgrids, with each subgrid assigned to a
processor.  Data on each subgrid are evolved independently during a
timestep and data on subgrid boundaries are exchanged when required
through message-passing calls.  This results in a highly scalable code
that exhibits good speedup over the full range of practically
realizable subgrids.

\subsection{Transport and Source Steps}

The transport step, as its name implies, handles the transport terms in the
continuity, energy, and momentum equations,
\begin{equation}
{\left[\partial_t {\cal{X}}\right]}_{transport} = -{1 \over \sqrt{\gamma}}\,
 \partial_i\left[\sqrt{\gamma}\,{\cal{X}}\,V^i\right]  \, , \, 
\left({\cal{X}} \equiv D, E, S_j\right).
\end{equation}
Other than the modified definition of the inertia, these
expressions are completely analogous to those in HSWb. 

The source step handles the remaining terms in the energy and momentum
equations.  The energy terms 
\begin{equation}
{\left[\partial_t E\right]}_{source} = -P\,\partial_{t} W - 
{P\over\sqrt{\gamma}}\,\partial_{i}\left[\sqrt{\gamma}\,W\,V^i\right] 
\end{equation}
are evolved as described in HSWb.  The momentum source terms contain 
new elements, namely
\begin{equation}
{\left[\partial_t S_j\right]}_{source} = 
-\alpha\,\partial_t\left[b_j\,b^t\right]+
  {\partial_i \left[\sqrt{\gamma}\,\alpha\,b_j\,b^i\right] \over 
  \sqrt{\gamma}}
  -\left[{S_\epsilon\,S_\mu \over S^t}-
  \alpha\,b_\mu\,b_\epsilon\right]\,
  {\partial_j g^{\mu\,\epsilon} \over 2}-
  \alpha\,\partial_j\left[P+{{\|b\|}^2 \over 2}\right].
\end{equation}
In the following expressions, superscript indices for the 
current time step are
understood.  The spatial differencing of some value $f$ is shown 
in compact form through the shift operator $D_i(f)$; 
any averaging of zone- and face-centered quantities that may be
required to properly center calculations is done as described in HSWb.

The time derivative of the magnetic fields is
evaluated from the newly-obtained values from the induction step,
and stored values from the previous time step,
\begin{equation}\label{src.6}
{{S}_{j}}^{n+1} = {S}_{j} + \alpha\,\left[
 {\left(b_j\,b^t\right)}^{(n+1)}-{\left(b_j\,b^t\right)}\right] .
\end{equation}
The magnetic gradient term discretizes readily,
\begin{equation}\label{tran.2}
{S_j}^{n+1} = {{S}_{j}}+{\Delta\,t\,\over \sqrt{\gamma}}\,
  {D_i \left(\sqrt{\gamma}\,\alpha\,b_j\,b^i\right)
  \over \Delta\,x^i} ,
\end{equation}
as does the pressure gradient term,
\begin{equation}\label{src.2}
{{S}_{j}}^{n+1} =  {S}_{j}-\Delta\,t\,\alpha \,
  {D_j\left(P+{\|b\|}^2/2\right) \over \Delta\,x^j}.
\end{equation}
Finally, the accelerations due to gradients in the metric,
\begin{equation}\label{src.4}
{{S}_{j}}^{n+1} = {S}_{j} - {\Delta\,t \over 2}\,
\left[{S_\epsilon\,S_\mu \over S^t}-
  \alpha\,b_\mu\,b_\epsilon\right]\,
  \partial_j\,g^{\mu\,\epsilon}.
\end{equation}
are computed using $S^t$ and $S_t$ obtained from $S_j$ using momentum
normalization (\ref{momnorm});  the metric derivatives,
$\partial_j\,g^{\mu\,\epsilon}$, are evaluated analytically over
the whole grid and stored in memory at the beginning of the
simulation.

In the source step, after the pressure gradient and acceleration terms have been
computed, the transport velocity $V^i= S^i/S^t$, and the Lorentz factor
$W=\alpha\,U^{t}$ are obtained using velocity normalization.  To guarantee that
the normalization is well-behaved in evacuated zones, we work with the momenta
$S_i$ rather than applying the normalization condition directly to the
velocities.  Although the two normalizations are analytically equivalent, 
use of the momentum protects against numerical rounding errors that cause
spurious floating point exceptions when extremely tenuous material is moving at
velocities close to that of light.

The source step also includes an artificial viscosity in both the
energy and momentum equations to provide a mechanism to increase the
entropy of the fluid in shocks.  The artificial viscosity is unchanged
from HSWb, except for modifying the definition of the inertia to
include the magnetic energy.

\subsection{The Induction Step: A General Relativistic Form of 
the MOCCT Algorithm}

The CT framework of EH88 is based on two basic ideas:  the use of
$F_{\mu\nu}$ as the fundamental variable so that the induction equation
reduces to simple partial derivatives, and a staggered-grid centering
of terms so that the numerical divergence of the CT magnetic field is
constrained to be zero.  The CT magnetic field component ${\cal{B}}^j$
is located on the $j$ face of a zone (a cube in three dimensions), and
the CT EMFs, ${\cal{E}}^i$, that are differenced to evolve ${\cal{B}}^j$
are located on the zone edges that define the $j$ face.

The CT formalism imposes no stringent conditions on exactly how the
EMFs are computed.  EH88 noted that some of the magnetic field
variables appear in the discretrized equations for the EMFs in the
guise of transport terms, requiring that they be upwinded for
stability, while others appeared as shear terms, apparently requiring
no special treatment.  EH88 used simple space-centered differencing for
those terms, but Stone \& Norman (1992) found that this prescription is
inadequate.  The results from simple Alfv\'en pulse propagation tests
where the Alfv\'enic velocity exceeds that of the background fluid found
substantial undamped dispersion error, resulting in excessive
zone-to-zone noise.

Stone \& Norman (1992) proposed to calculate the CT EMFs by
solving a one dimensional analytic linear Alfv\'en wave equation to
obtain improved values of ${\cal{B}}^j$ and $V^j$ in the construction
of the EMFs.  This approach became known as the Method of Characteristics
Constrained Transport (MOCCT).  Since then several variants have been
proposed (e.g. Hawley \& Stone 1995; Clarke 1996) to further improve
the computation of Newtonian MHD.  While the traditional approach to
MoC does not carry over readily to GR MHD, it is nevertheless possible
to address the problem by using the same motivation that underlies
MOCCT.  Instead of using an analytic Alfv\'en wave solution we use a
numerical solution of the simplified one-dimensional induction and
momentum equations to obtain the improved EMFs.

Our implementation of CT for the GRMHD code is accomplished as
follows.  Following the example of Hawley \& Stone (1995), the
starting point for constructing, for instance, the $z$-EMF,
\begin{equation}
{\cal{E}}^z = V^x\,{\cal{B}}^y - V^y\,{\cal{B}}^x,
\end{equation}
lies in obtaining the discretized solutions to the 1D incompressible
MHD equations.  On a staggered mesh, ${\cal{E}}^z$ is located on zone
edges in such a way that it is constructed from $x$- and $y$-components
relocated in the $y$- and $x$-directions respectively.  To take a
specific example, we would like obtain an edge-centered estimate
${{\cal{B}}^y}^*$ to construct the above EMF.  We do this by solving
the 1D linearized induction equation in the $x$-direction (i.e.
$\partial_y \equiv \partial_z \equiv 0$) for small transverse ($y$)
perturbations.  This simplified induction equation has the form,
\begin{eqnarray}
\label{alfdiv}
\partial_x\,{\cal{B}}^x & = & 0 , \\
\label{alfind}
\partial_t\,{\cal{B}}^y & = & 
{\cal{B}}^x\,\partial_x\,\left(V^y\right)- 
V^x\,\,\partial_x\left({\cal{B}}^y \right).
\end{eqnarray}
By applying the usual differencing we obtain for the left-hand side,
\begin{equation}
\partial_t\,{\cal{B}}^y = {{{\cal{B}}^y }^*-\overline{{\cal{B}}^y } \over 
\Delta t} ,
\end{equation}
where the overbar denotes the original EH88 upwinding. 
We can now readily obtain estimates of the starred quantity,
\begin{equation}
{{\cal{B}}^y}^* =\overline{{\cal{B}}^y} + {1 \over 2}\,
 {\Delta t \over \Delta x}\,\left[-V^x\,D_x({\cal{B}}^y)
  +{\cal{B}}^x D_x(V^y) \right] ,
\end{equation}
where the factor of $1/2$ ensures proper time-centering of the starred
quantity (the starred values are centered between the $n$ and
$n+1$ time levels).  A similar approach can now be used to generate
${{\cal{B}}^x}^*$, and produce an improved estimate of ${\cal{E}}^z$.
The procedure is just as easily applied to the other two EMFs to
complete the process for all CT variables. Note that because of the
fully covariant nature of the CT equations, this process applies to any
coordinate system.

For consistency, we also need to evolve the linearized momentum
equation for the transverse velocities.  Care must be taken in deriving
these equations since, in contrast to the CT induction equation, metric
terms are explicitly involved in the expressions.  Continuing with the
example for ${\cal{E}}^z$, and regarding the $x$ direction to be either
$r$ or $\theta$, and $y$ a direction transverse to this, we obtain
\begin{equation}\label{alfvel}
\left(\partial_t + \xi\,V^x\,\partial_x\right)\,V^y - 
\kappa\,\,{\cal{B}}^x\partial_x\,{\cal{B}}^y \approx 0.
\end{equation}
where,
\begin{eqnarray}
\xi & = & {\rho\,h\,W^2 - {\|b\|}^2 \over \rho\,h\,W^2 + {\|b\|}^2 },\\
\kappa & = & {\alpha^2 \over
{\left(W\,\sqrt{4\,\pi\,\gamma} \right)}^2 
\left(\rho\,h\,W^2 + {\|b\|}^2 \right) }.
\end{eqnarray}
(The details are found in Appendix A.) Note that the above equations
are similar to the Newtonian results, but with 
the relativistic factors $\xi$ and $\kappa$.  By
simply applying the usual differencing, however, we readily obtain
expressions for an improved estimate of the transverse velocity,
\begin{equation}
\partial_t\,V^y = {{V^y}^*-\overline{V^y} \over \Delta t}
\end{equation}
where the overbar denotes the edge-centered average, from which,
\begin{equation}
{V^y}^* = \overline{V^y} + {1 \over 2}\,
 {\Delta t \over \Delta x}\,\left[-\xi\,V^x\,D_x (V^y) + 
  \kappa\,{\cal{B}}^x\,D_x\,{\cal{B}}^y \right].
\end{equation}

In the Kerr metric, the calculation in the $\phi$ direction is somewhat
different due to the $g^{t\phi}$ metric term.  In this case
we obtain 
\begin{equation}\label{alfvel_phi}
\left[\partial_t + \left(\xi\,V^\phi+\lambda\right)\,\partial_\phi\right]\,V^y - 
\kappa\,\,{\cal{B}}^\phi\partial_\phi\,{\cal{B}}^y \approx 0.
\end{equation}
where $\xi$ and $\kappa$ are unchanged and
\begin{equation}
\lambda = {2\,{\|b\|}^2\,g^{t \phi} \over g^{tt}
\left(\rho\,h\,W^2 + {\|b\|}^2 \right)}.
\end{equation}

Combined with the starred quantities for the CT variables, we now have a
complete set of variables from which to construct the EMFs.

\section{1D Test Cases - Minkowski Spacetime}

The first series of tests is designed to verify code performance in the
special relativistic limit.  For this purpose we use a one-dimensional
version of the code and the Minkowski metric.  The tests include
Alfv\'en wave propagation, fast and slow magnetosonic shocks,
rarefaction waves, and both relativistic and non-relativistic shock
tubes.  Many of these tests have appeared previously in the literature
in one form or another.  In particular, Komissarov (1999) (hereafter
referred to as K99) presented a series of challenging test problems for
a special relativistic MHD Godunov scheme, and Stone {\it{et al.}}
(1992) presented a test suite for Newtonian MHD codes.  Our
presentation will conform to the formats of these published results to
facilitate comparison.

\subsection{Alfv\'en Wave Propagation}

The propagation of a linear Alfv\'en wave is simple in conception, but,
as demonstrated by Stone \& Norman (1992), can be quite revealing as a
numerical test.  The special relativistic Alfv\'en wave includes the
displacement current normally neglected in Newtonian MHD.  This is what
limits wave speeds to less than $c$.

\begin{figure}[ht]
     \epsscale{0.5}
     \plotone{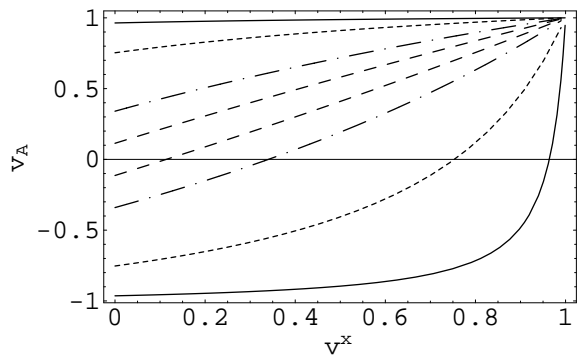}
     \caption{\label{Valfven} 
     Plot of ${v_A}^{(\pm)}$ for $\beta={10}^{-3}$ (solid line), 
     $\beta={10}^{-2}$ (short dashes),
     $\beta={10}^{-1}$ (dash-dot), $\beta=1$ (long dashes).}
\end{figure}

We consider a fixed background magnetic field ${\cal{B}}^x$ in
Minkowski spacetime with constant fluid velocity $V^x$ and a small
transverse perturbation with velocity $V^y$ and CT magnetic field
${\cal{B}}^y$.  The induction and momentum equations for the perturbed
quantities have the form given in equations (\ref{alfind}) and
(\ref{alfvel}).  Equation (\ref{alfvel}) can be combined with the
induction equation (\ref{alfind}) by taking partial derivatives
[$\left(\partial_t + V^x\,\partial_x\right)$ for the former and
$\partial_x$ for the latter] to get an equation for $V^y$,
\begin{equation}\label{alfven.4}
\left[\partial^2_t+\left(1+\xi\right)\,V^x\,\partial_t\,\partial_x+
\left(\xi\,{(V^x)}^2-\kappa\,{({\cal{B}}^x)}^2\right)\,
\partial_x^2\right]\,V^y = 0 
\end{equation} 
which describes the propagation of Alfv\'en waves with speeds
\begin{equation}\label{alfven.5} 
{v_A}^{(\pm)} = {V^x \pm \eta\,\sqrt{\eta^2 +W^{-2}} \over 1+\eta^2} 
\rightarrow \pm {\sqrt{{\|b\|}^2 \over \rho\,h+{\|b\|}^2}} \, 
(for\,V^x=0) ,
\end{equation} 
where $\eta^2 = {\|b\|}^2/(\rho\,h\,W^2)$.  (The equation for
${\cal{B}}^y$ can be obtained in a similar manner.) Figure
\ref{Valfven} plots ${v_A}^{(\pm)}$ as a function of the background
velocity of the fluid ($V^x$) for four values of $\beta$ and
illustrates several key properties of the wave speeds:
(1) for $V^x=0$, waves move with equal and opposite speed 
(for all $\beta$); 
(2) ${v_A}^{(\pm)}$ is constrained to the upper bound of $c \equiv 1$;
(3) for a given $\beta$, there is a $V^x_0$ that yields a stationary wave,
${v_A}^{(-)}=0$;
(4) for $V^x > V^x_0$, both waves move in the same direction, with
${v_A}^{(-)}$ lagging behind the main fluid flow, and ${v_A}^{(+)}$ leading;
(5) as $V^x \rightarrow c$, all wave speeds converge, 
${v_A}^{(\pm)}\rightarrow c$.

We describe four tests, listed in Table \ref{1Dpulse}, carried out in a
periodic box of width $x_{max}-x_{min}=3.0$.  In each case the fluid
has constant density $\rho = 1$ and energy density $\epsilon =
{10}^{-2}$.  Individual tests are distinguished by the amplitude of the
$x$-component of the CT magnetic field ${\cal{B}}^x$, given here in
terms of $\beta=\sqrt{2\,P/{\|b\|}^2}$, and the amplitude of the
background $x$-component fluid velocity $V^x$.  The waves are
initialized with a square pulse in the transverse velocity $V^y$
with amplitude $A_0=10^{-3}$, namely
\begin{equation}\label{Vytweak} 
V^y = \cases{ 
    A_0 & for $1.0 < x < 1.5$ \cr 
   -A_0 & for $1.5 \le x < 2.0$}. 
\end{equation}
The transverse magnetic field is initially zero, ${\cal{B}}^y=0$.

\begin{figure}[ht]
     \epsscale{0.2}
     \plotone{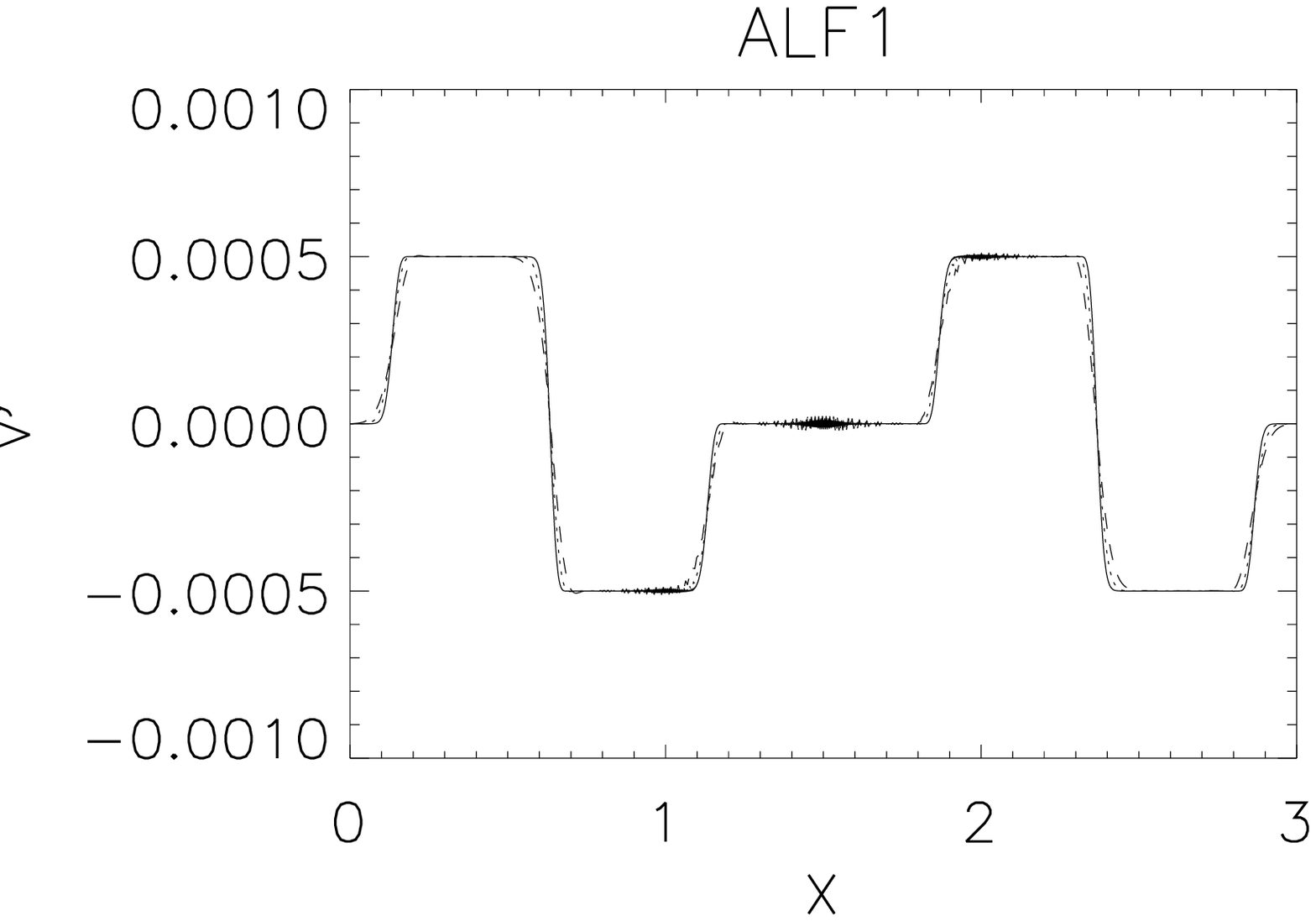}
     \plotone{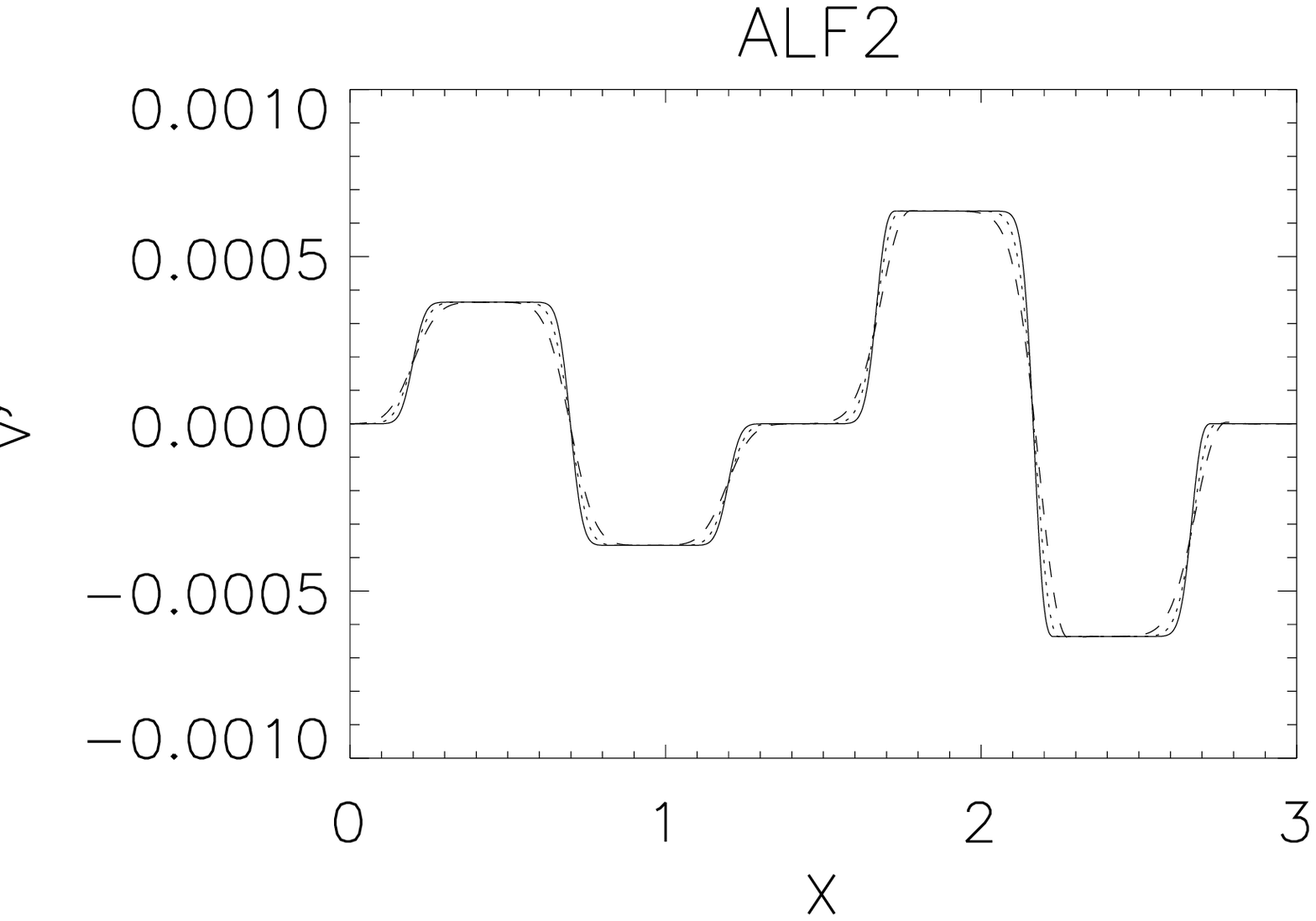}
     \plotone{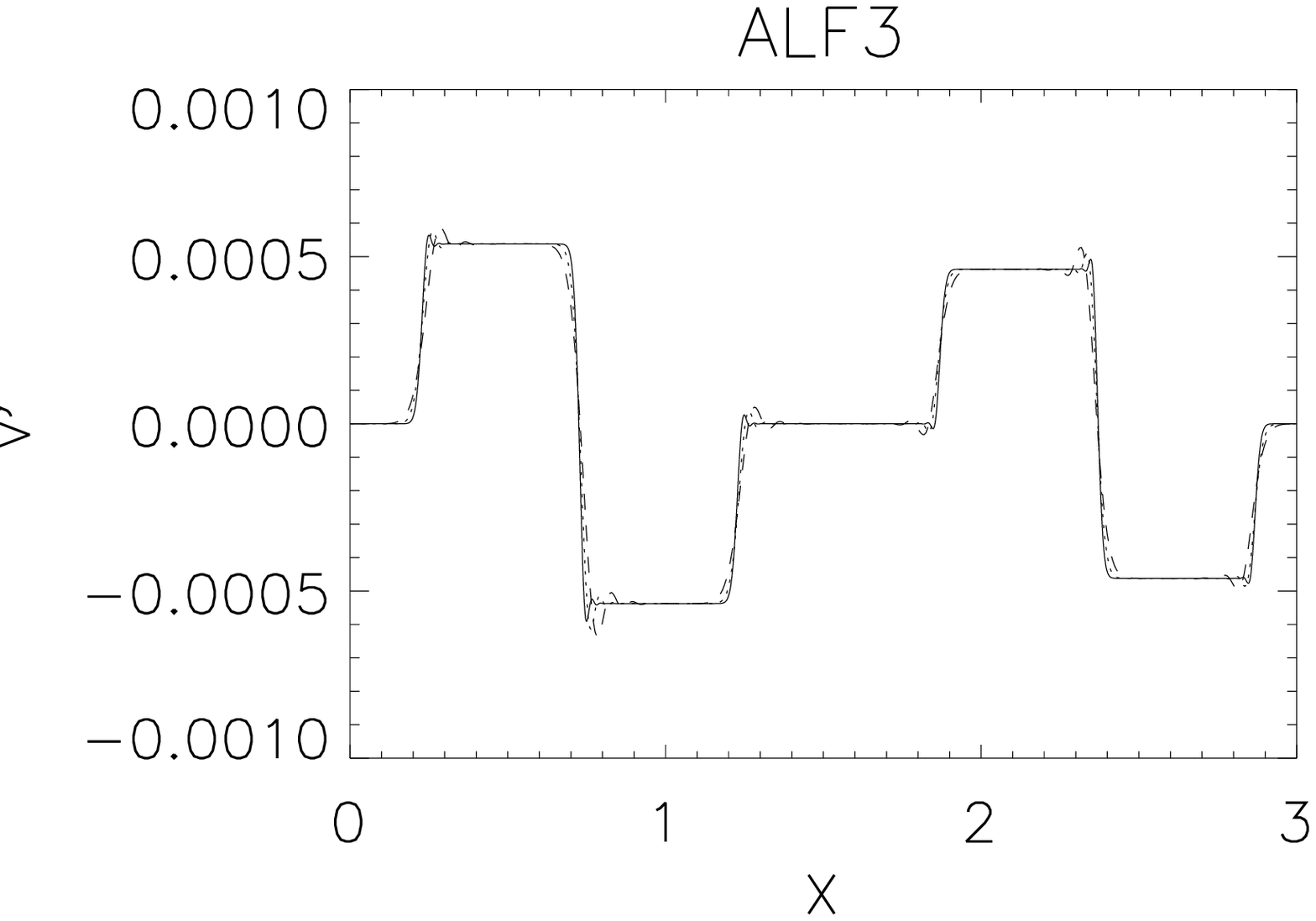}
     \plotone{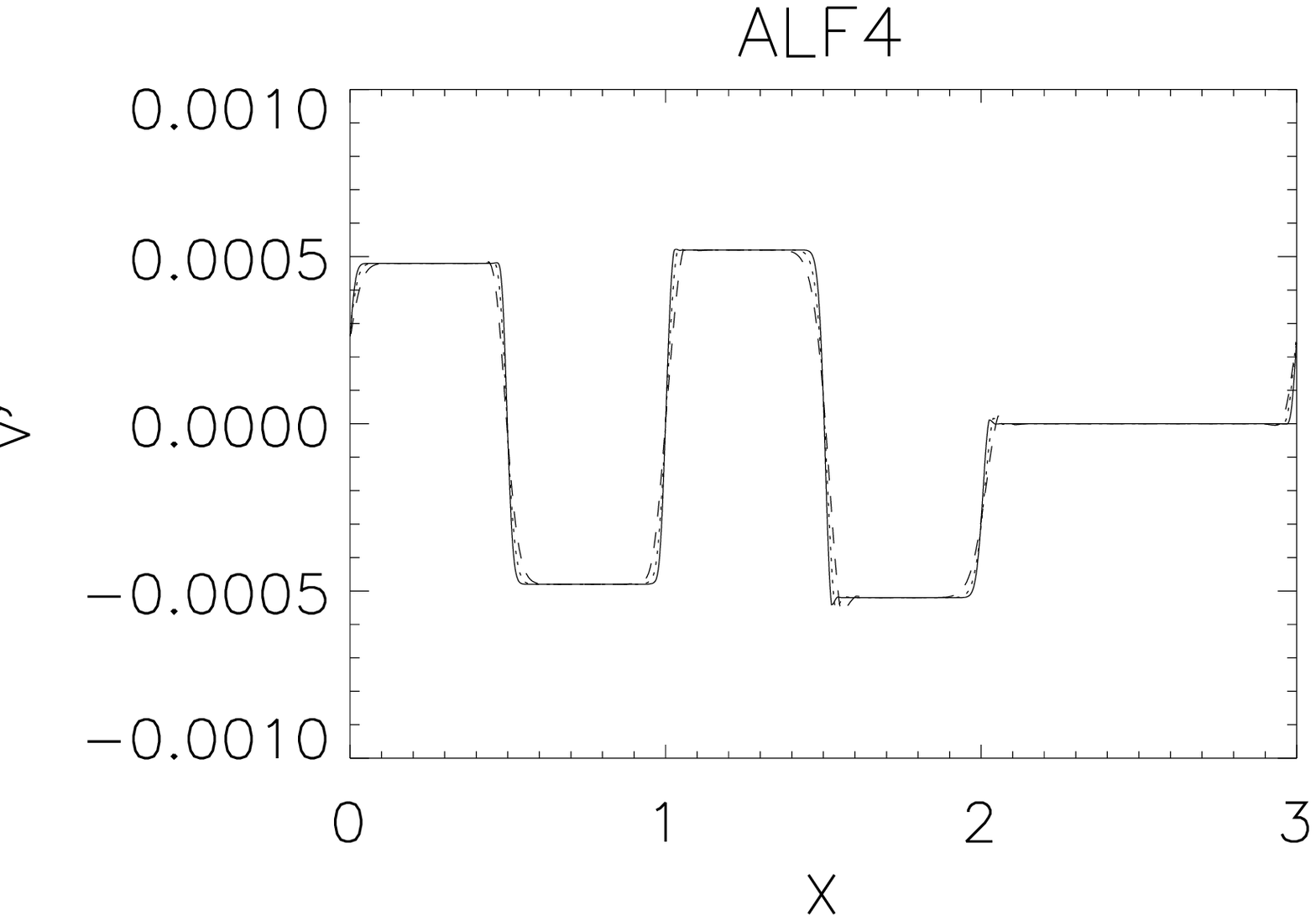}
     \caption{\label{V1pulse} 
     Transverse velocity $V^y$ at time $t_{final}$ for Alfv\'en wave
     tests (from left to right) ALF1, ALF2, ALF3, and ALF4.  Three
     different resolutions are overlaid:  512, 1024, and 2048
     zones.} 
\end{figure}

Table \ref{1Dpulse} summarizes the results and Figure
\ref{V1pulse} overlays $V^y$ for each test at time $t_{final}$ at three
grid resolutions, $512$, $1024$, and $2048$ zones.  In test ALF2 the
fast and slow pulses have wrapped around and re-entered the grid from
the left, the fast pulse having done so twice.  In ALF4, the fast pulse
has also wrapped around while the slow pulse is stationary.  (In this
linear regime the pulses overlap cleanly provided that the spatial
resolution is sufficiently high to minimize losses due to numerical
diffusion.) Measuring from the center of each pulse (originally at
$x=1.5$) we verify that the observed pulse speeds agree with
(\ref{alfven.5}).  With a stationary background, ALF1, the pulses
separate from the initial perturbation with equal and opposite
velocities and also with equal amplitudes such that $A_0 = A^{+} +
A^{-}$.  In a moving background, the splitting of the velocity pulses
is no longer
symmetrical, as is especially obvious in ALF2 (although $A_0 = A^{+} +
A^{-}$ still holds).  The slow velocity pulse has a greater amplitude than the
fast pulse.  We find that the ratio of the velocity amplitudes scales inversely
with their Lorentz factors,
\begin{equation}\label{alfgam}
{A^{+} \over A^{-}} = {W({v_A}^{-}) \over W({v_A}^{+})}.
\end{equation}
It is easy to verify that this holds for each model in Table \ref{1Dpulse}.
The pulses in the magnetic field variable, ${\cal B}^y$, on the other hand, 
are equal in magnitude and opposite in phase for all $V^x$. The numerical
values given in Table \ref{1Dpulse} can be verified by direct computation
using results presented in Appendix B.

\begin{table}[ht]
\caption{\label{1Dpulse} Results of 1D Alfv\'en Pulse Tests.}
\begin{tabular}{lrrrrrrrrrr}
 & & & & & & & & & & \\
\hline
Model & $V^x$ & $\beta$ & ${v_A}^{+}$ & $A^{+}$ {\tiny{$(\times {10}^{-4}$)}} & $W^{+}$ 
                        & ${v_A}^{-}$ & $A^{-}$ {\tiny{$(\times {10}^{-4}$)}} & $W^{-}$
      & $t_{final}$ & $\|{\cal B}^y\|$ {\tiny{$(\times {10}^{-3}$)}}\\
\hline
\hline
ALF1 &  0.000 & 0.001 & +0.96 & 5.00 & 3.76 & -0.96 & 5.00 & 3.76  & 0.9 & 6.714\\
ALF2 &  0.800 & 0.100 & +0.90 & 3.64 & 2.26 & +0.63 & 6.36 & 1.29  & 5.8 & 4.888\\
ALF3 &  0.100 & 0.010 & +0.79 & 4.62 & 1.64 & -0.71 & 5.38 & 1.41  & 1.1 & 2.729\\
ALF4 &  0.200 & 0.315 & +0.38 & 4.80 & 1.08 &  0.00 & 5.20 & 1.00  & 6.2 & 1.897\\
\hline
\end{tabular}
\end{table}

\subsection{Magnetosonic Shock and Rarefaction Tests}

Shock waves and rarefactions test a code's ability to respond to
discontinuities and to maintain the jump conditions that correspond to
the conservation properties of the flow.  In this section we consider a
number of one-dimensional shock tests, both relativistic and
non-relativistic, that involve nonzero transverse magnetic fields.

We begin with relativistic magnetosonic shock wave tests; the initial
left ($x < 0.0$) and right ($x \ge 0.0$) states are given in Table
\ref{1DSHK}.  In each case we verified that these states satisfy the
relativistic shock invariants (see Appendix C).  In this and in
subsequent tables we note those tests that were described in
K99.\footnote{K99 magnetic field variables ${\cal{B}}^i_{K99}$ are
related to our CT-variables by a normalization factor, ${\cal{B}}^i =
\sqrt{4\,\pi}\,{\cal{B}}^i_{K99}$.  Also, a revised version of K99
Table 2 was given by Komissarov (2002).}

\begin{table}[ht]
\caption{\label{1DSHK} Initial States for 1D Magnetosonic Shock Tests.}
\begin{tabular}{lllll}
 & & & & \\
\hline
Test & Left State & Right State & Grid & $t_{final}$\\
\hline
\hline
Slow Shock & $U^i=(1.53,0.0,0.0)$ & $U^i=(0.9570,-0.6820,0.0)$ 
           & $512$ & 2.0  \\
($V=0.5$)  & $V^i=(0.838,0.0,0.0)$ & $V^i=(0.620,-0.442,0.0)$ & &\\
(K99)      & ${\cal{B}}^i=\sqrt{4\,\pi}(10.0,18.28,0.0)$ 
           & ${\cal{B}}^i=\sqrt{4\,\pi}(10.0,14.49,0.0)$ & & \\
           & $P=10.0$ $\rho=1.0$ & $P=55.33$ $\rho=3.322$ \\
 & & & & \\
Fast Shock I & $U^i=(1.780,0.114,0.0)$ & $U^i=(0.922,0.403,0.0)$ 
           & $1024$ & 2.5 \\
($V=0.0$)  & $V^i=(0.870,0.056,0.0)$ & $V^i=(0.650,0.284,0.0)$ & & \\
($W \approx 2.0$) & ${\cal{B}}^i=\sqrt{4\,\pi}(3.33,2.50,0.0)$
           & ${\cal{B}}^i=\sqrt{4\,\pi}(3.33,4.52,0.0)$ & & \\
           & $P=2.015$ $\rho=1.406$ & $P=5.135$ $\rho=2.714$ \\
 & & & & \\
Fast Shock II & $U^i=(1.780,0.114,0.0)$ & $U^i=(1.479,0.280,0.0)$ 
           & $1024$ & 2.5 \\
($V=0.2$)  & $V^i=(0.870,0.056,0.0)$ & $V^i=(0.818,0.155,0.0)$ & & \\
($W \approx 2.0$) & ${\cal{B}}^i=\sqrt{4\,\pi}(3.33,2.50,0.0)$
           & ${\cal{B}}^i=\sqrt{4\,\pi}(3.33,3.25,0.0)$ & & \\
           & $P=2.015$ $\rho=1.406$ & $P=2.655$ $\rho=1.725$ \\
 & & & & \\
Fast Shock III & $U^i=(3.649,0.114,0.0)$ & $U^i=(0.715,0.231,0.0)$ 
           & $128$ & 2.5 \\
($V=0.2$)  & $V^i=(0.964,0.030,0.0)$ & $V^i=(0.542,0.185,0.0)$ & & \\
($W \approx 3.8$) & ${\cal{B}}^i=\sqrt{4\,\pi}(3.33,2.50,0.0)$
           & ${\cal{B}}^i=\sqrt{4\,\pi}(3.33,6.52,0.0)$ & & \\
           & $P=2.015$ $\rho=1.406$ & $P=34.99$ $\rho=8.742$ \\
\hline
\end{tabular}
\end{table}

\begin{figure}[ht]
     \epsscale{0.25}
     \plotone{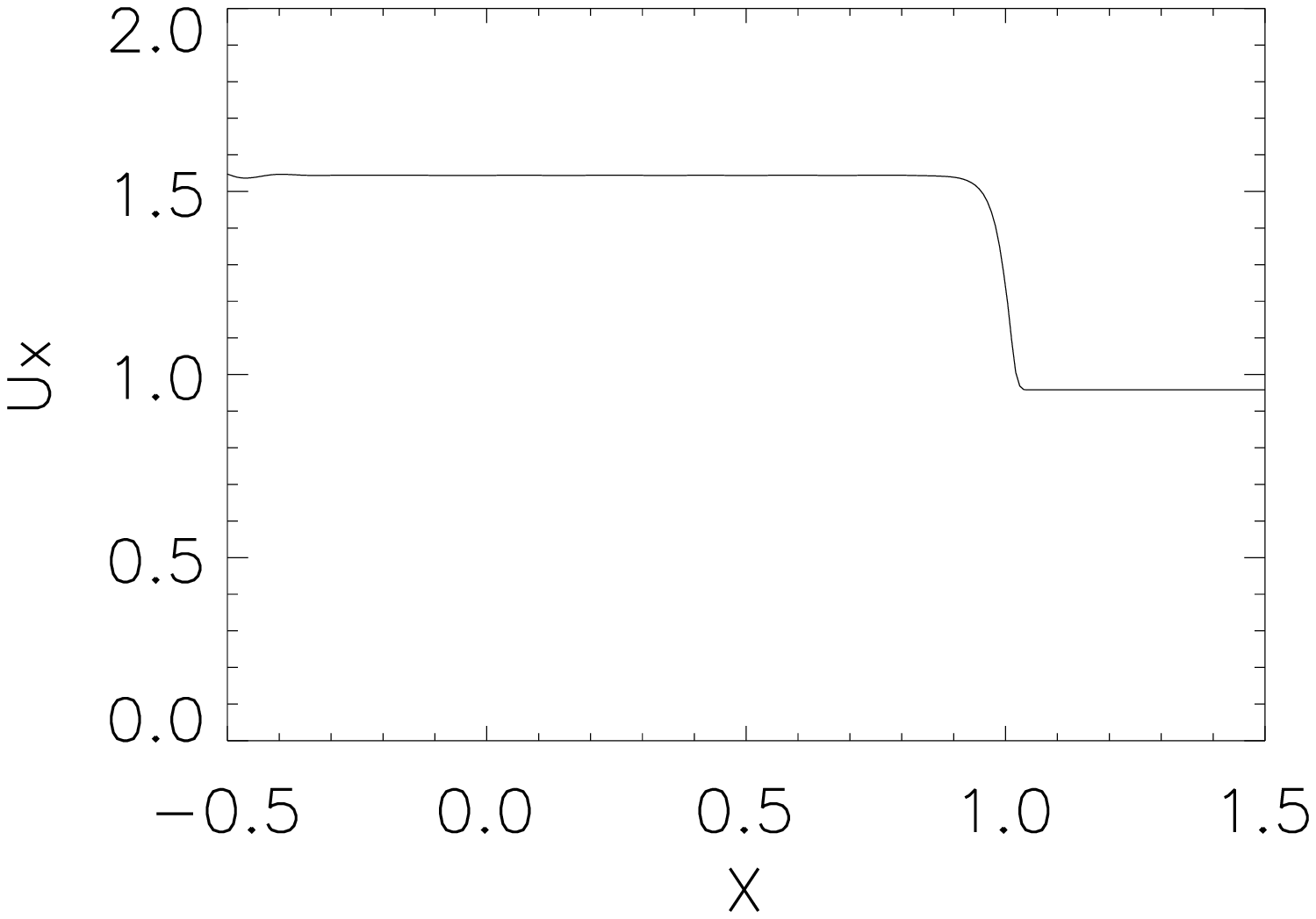}
     \plotone{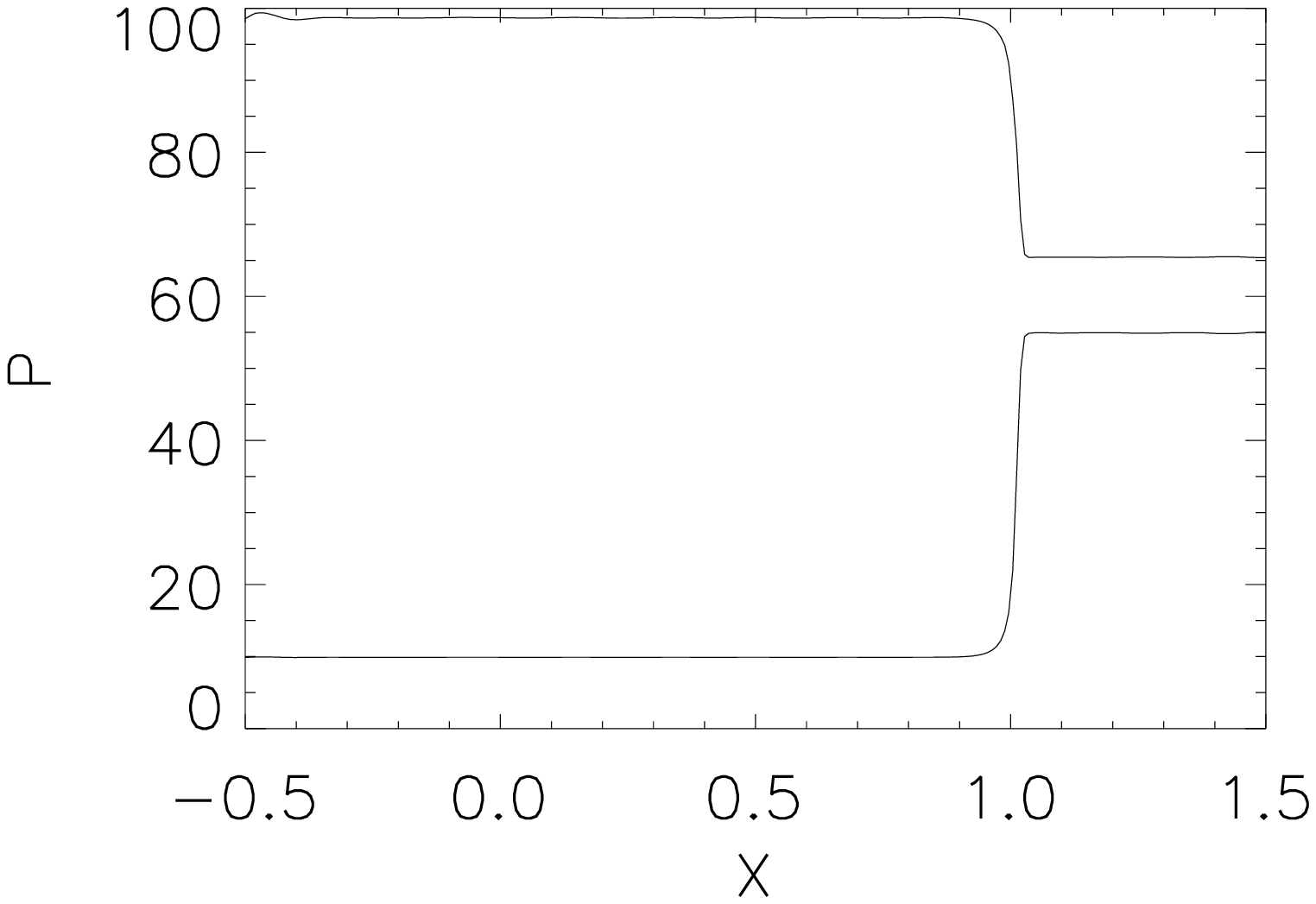}
     \plotone{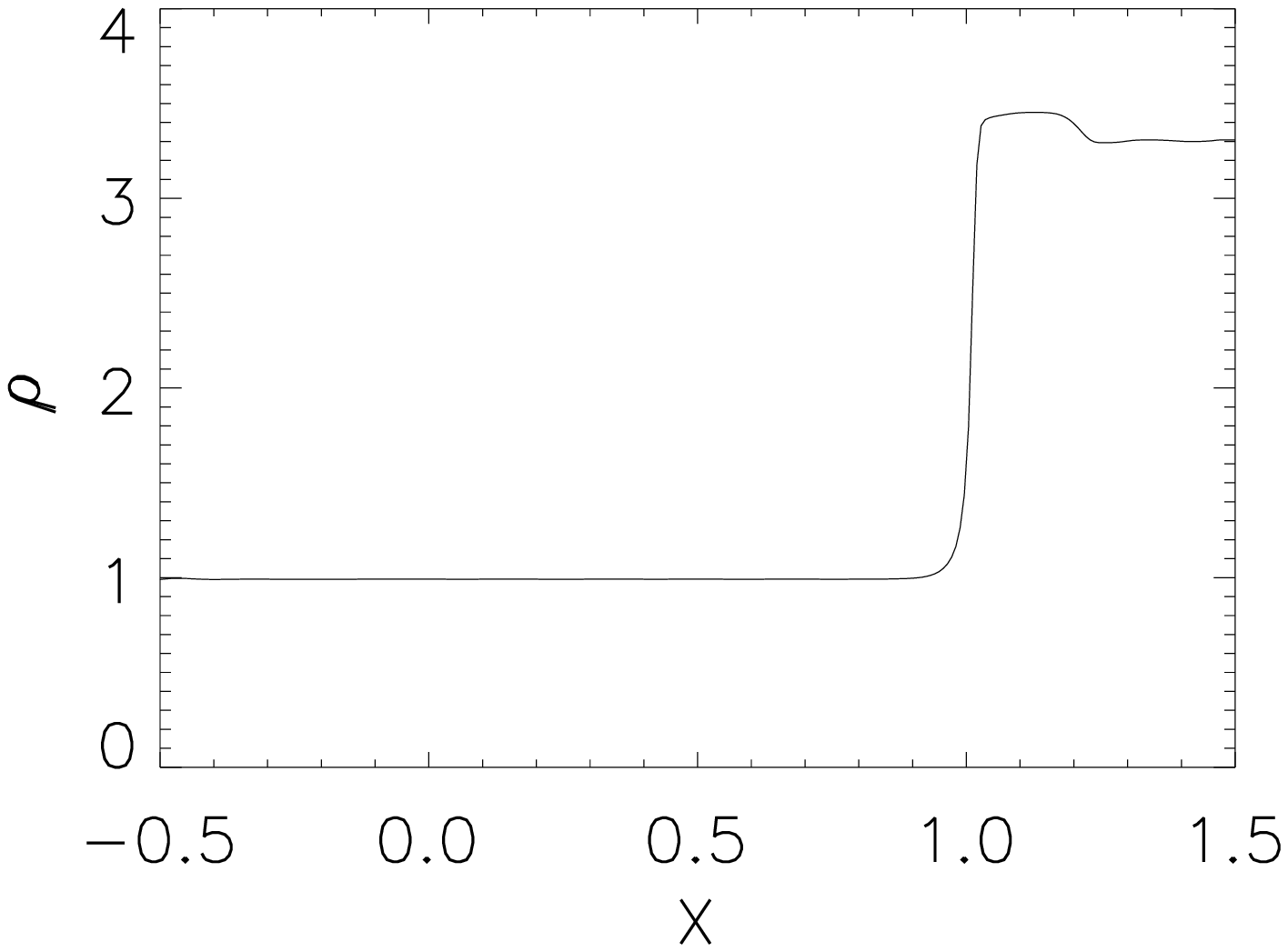}
     \caption{\label{SLOWSH} 
     Plot of $U^x$ (left), $P_{gas}$ (bottom curve, center) and $P_{mag}$ 
     (top curve, center), and $\rho$ (right) at $t = 2.00$ for the K99 Slow 
     Magnetosonic Shock.
     }
\end{figure}

The first test, the slow magnetosonic shock, is shown in Figure
\ref{SLOWSH}, which plots $U^x$, $\rho$, gas pressure $P$, and magnetic
pressure $P_{mag}$ at $t=2.0$, by which point the shock front has moved
from $x=0.0$ to $x=0.5$.  These graphs are quite similar to the
results of K99.  There is a small step in density in the high-density
medium to the right of the moving shock.  When we repeat our tests at
lower resolution, our step feature resembles the ``wiggle'' visible in
K99.  The origin of this small step may be due to a small discrepancies
in the initial state (i.e.  the data in K99, which we use here, differ
slightly from what we compute using the method outlined in Appendix
C).

\begin{figure}[ht]
     \epsscale{0.45}
     \plotone{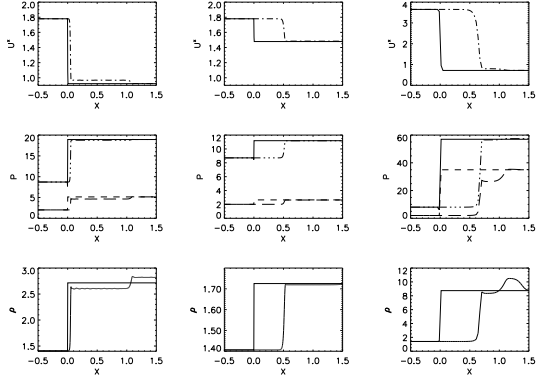}
     \caption{\label{FASTS1} 
     Plot of $U^x$, $P_{gas}$ and $P_{mag}$, and $\rho$ at $t = 2.50$, 
     for the Fast Magnetosonic Shock I (left column);
     the Fast Magnetosonic Shock II (center column); 
     and the Fast Magnetosonic Shock III (right column).}
\end{figure}

Figure \ref{FASTS1} shows plots of $U^x$, $\rho$, gas pressure $P$, and
magnetic pressure $P_{mag}$ for the fast magnetosonic shocks at
$t=2.5$.  In the case of the stationary shock (I), we see that
artificial viscosity has slightly altered the shock front, and that
there is an overvalue in the left state $U^x$, and undervalue in $\rho$
and $P_{gas}$.  In the case of moving shock (II), the shock front has
traveled to $x=0.5$, as expected for the given shock speed and the
value of $t_{final}$, and the left and right states agree with the
analytic values.  In the case of the $W \approx 3.8$ shock (III),
a small mismatch has developed in the post-shock gas pressure near the
shock edge.  In addition, the speed of the shock slightly exceeds the
expected value $0.2$.  As discussed by HSWb and studied in detail by
Norman \& Winkler (1986), the artificial viscosity used in the code to
increase the entropy of the fluid through a shock produces sufficient
heating for $W < 2$, but increasingly underestimates the heating
as $W$ increases.  The results of this shock test are consistent
with these earlier findings.

\begin{figure}[ht]
     \epsscale{0.45}
     \plotone{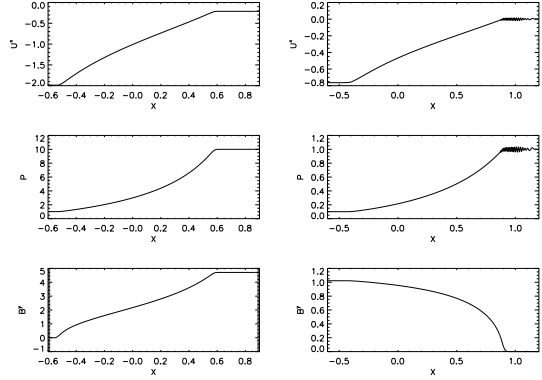}
     \caption{\label{RAREFAC} 
     Plot of $U^x$, $P_{gas}$, and $\rho$ at $t = 1.00$ in the
     K99 Fast Rarefaction Test (left column); and
     plot of $U^x$, $P_{gas}$, and $\rho$ at $t = 2.00$ in the
     K99 Slow Rarefaction Test (right column).}
\end{figure}

We next consider a pair of challenging rarefaction tests designed by
K99.  The initial left and right states of these tests are given in
Table \ref{1DRAR}.  Figure \ref{RAREFAC} shows a plot of $U^x$, gas
pressure $P$ and density $\rho$ for the switch-off fast rarefaction at
$t=1.0$ and the switch-on slow rarefaction at $t=2.0$.  Our results
generally agree with those of K99.  The fast rarefaction extends from
$x=-0.5$ to $x=+0.6$ and the slow rarefaction extends from $x=-0.5$ to
$x=+0.9$.  There is high-frequency noise in the plot of $U^x$ and $P$
to the right of the slow rarefaction.  We attribute this noise to an
artifact of the routine that converts CT-variables to fluid-frame
magnetic field.  This test is particularly sensitive to the averaging
techniques that are an unavoidable part of staggered-mesh
discretizations.  This noise is reduced at lower grid resolution, where
increased numerical diffusion acts to damp out the oscillations.

\begin{table}[ht]
\caption{\label{1DRAR} Initial States for 1D Rarefaction Tests.}
\begin{tabular}{lllll}
 & & & & \\
\hline
Test & Left State & Right State & Grid & $t_{final}$\\
\hline
\hline
Switch-off Fast&  $U^i=(-2.0,0.0,0.0)$ & $U^i=(-0.212,-0.590,0.0)$ 
           & $2048$ & 1.0\\
Rarefaction& $V^i=(-0.894,0.0,0.0)$ & $V^i=(-.180,0.500,0.0)$ & &\\
(K99)      & ${\cal{B}}^i=\sqrt{4\,\pi}(2.0,0.0,0.0)$ 
           & ${\cal{B}}^i=\sqrt{4\,\pi}(2.0,4.71,0.0)$ & & \\
           & $P=1.0$ $\rho=0.1$ & $P=10.0$ $\rho=0.562$ \\
 & & & & \\
Switch-on Slow&  $U^i=(-0.765,-1.386,0.0)$ & $U^i=(0.0,0.0,0.0)$ 
           & $2048$ & 2.0\\
Rarefaction& $V^i=(-0.409,-0.740,0.0)$ & $V^i=(0.0,0.0,0.0)$ & &\\
(K99)      & ${\cal{B}}^i=\sqrt{4\,\pi}(1.0,1.022,0.0)$ 
           & ${\cal{B}}^i=\sqrt{4\,\pi}(1.0,0.0,0.0)$ & & \\
           & $P=0.1$ $\rho=1.78 \times {10}^{-3}$ & $P=1.0$ $\rho=0.01$\\
\hline
\end{tabular}
\end{table}

\begin{figure}[ht]
     \epsscale{0.5}
     \plotone{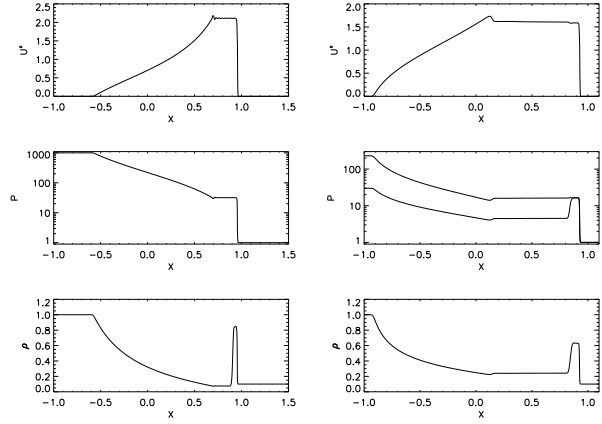}
     \caption{\label{SHOCK1} 
     Plot of $U^x$, $P_{gas}$ (and $P_{mag}$ at right), 
     and $\rho$ at $t = 1.00$ for K99 Shock Tube 1 (left column)
     and K99 Shock Tube 2 (right column).}
\end{figure}

\begin{figure}[hb]
     \epsscale{0.5}
     \plotone{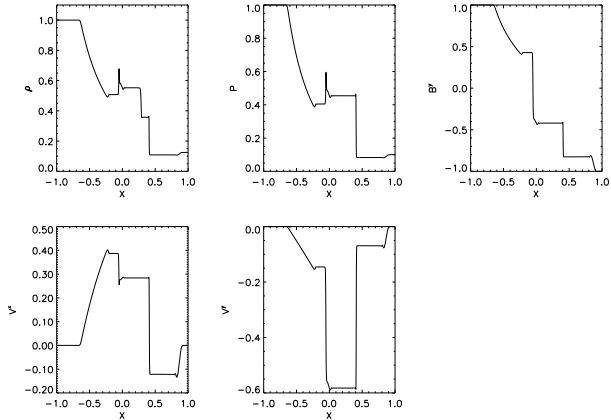}
     \caption{\label{BRIOWU} 
     Plot of $\rho$ (top row, left), $P_{gas}$ (top row, center), 
     ${\cal{B}}^y/\sqrt{4\,\pi}$ (top row, right), 
     $V^x$ (bottom row, left), 
     and $V^y$ (bottom row, center) at $t = 1.00$ for the
     Relativistic Brio and Wu Shock Tube.}
\end{figure}

\begin{figure}[ht]
     \epsscale{0.5}
     \plotone{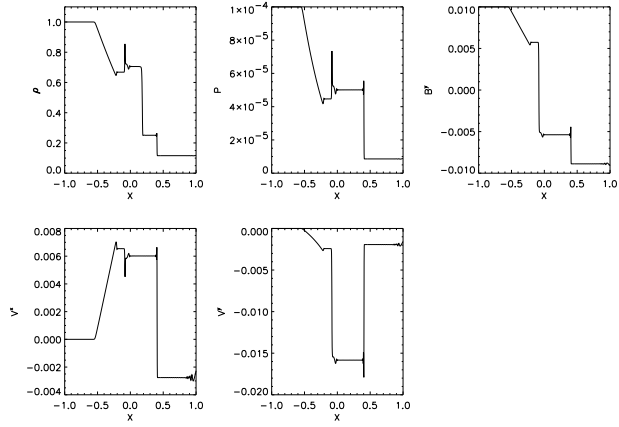}
     \caption{\label{BRIONR} 
     Plot of variables $\rho$ (top row, left), $P_{gas}$ (top row, center), 
     ${\cal{B}}^y/\sqrt{4\,\pi}$ (top row, right), 
     $V^x$ (bottom row, left), 
     and $V^y$ (bottom row, center) at $t = 30.00$, 
     in the $\gamma = 2$ Non-Relativistic Brio and Wu Shock Tube.}
\end{figure}

\begin{figure}[hb]
     \epsscale{0.5}
     \plotone{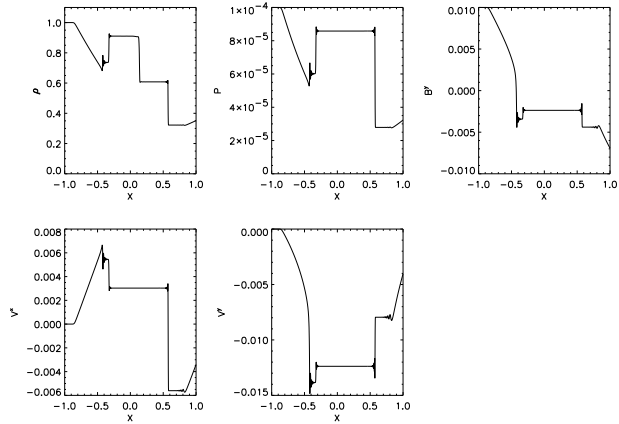}
     \caption{\label{RYUJON} 
     Plot of variables $\rho$ (top row, left), $P_{gas}$ (top row, center), 
     ${\cal{B}}^y/\sqrt{4\,\pi}$ (top row, right), $V^x$ (bottom row, left), 
     and $V^y$ (bottom row, center) at $t = 45.00$, Non-Relativistic Ryu and 
     Jones Shock Tube.}
\end{figure}

Next we consider a series of shock tube tests that combine strong shocks and
rarefactions of various types.  We ran two test problems from K99, the Brio \&
Wu (1988) shock tube test as described in Stone {\it{et al.}} (1992), an
alternate version of the Brio \& Wu shock tube using the same initial state but
with $\gamma=4/3$ instead of $\gamma = 2$ to ensure that the sound speed is not
relativistic, and a non-relativistic shock tube from Ryu \& Jones (1995).  In
the non-relativistic shock tubes the values of pressure and density are rescaled
to ensure that the sound speed is much less than $c$.  Table \ref{1DSHTU}
summarizes the various states for the shock tube tests, following the convention
of Stone {\it{et al.}} (1992).  Figures \ref{SHOCK1}, \ref{BRIOWU},
\ref{BRIONR}, and \ref{RYUJON} show our numerical results for the evolution of
the shock tubes at a grid resolution of $2048$ zones.  Some ringing is present
in the non-relativistic shock tubes, at the location of the compound wave and at
the slow shock in the density variable, for instance, whereas this
ringing is absent from the relativistic
shock tubes.  This appears to be a manifestation of decreased dispersion 
error at relativitistic velocities.

\begin{table}[ht]
\caption{\label{1DSHTU} Initial and Intermediate States for 
Shock Tube Tests.  ($\gamma=4/3$ unless indicated.)}
\begin{tabular}{llrrrrrrr}
 & & & & & & & & \\
\hline
Test & Variable & Left & $FR$ & $SC$ & $CD_l$ & $CD_r$ & $FR$ & Right \\
\hline
\hline
Shock Tube 1 & $\rho$& 1.00 & 0.08 &  ---  & 0.85 & 0.10 & --- & 0.10 \\
(K99)        & $P$   & 1000 & 31.8 &  ---  & --- & --- & --- & 1.00 \\
($v^y= {\cal{B}}^y=0$)
             & $v^x$ & 0.00 & 0.90 &  ---  & --- & --- & --- & 0.00 \\
\hline
Shock Tube 2 & $\rho$& 1.00 & 0.24 &  ---  & 0.63 & 0.10 & --- & 0.10 \\
(K99)        & $P$   & 30.0 & 4.50 &  ---  & 16.5 & 1.00 & --- & 1.00 \\
($v^y=0$)    & $v^x$ & 0.00 & 0.85 &  ---  & --- & --- & --- & 0.00 \\
 & ${\cal{B}}^y/\sqrt{4\,\pi}$& 20.0 & 9.14 &  ---  & --- & --- & --- & 0.00 \\
\hline
Brio and Wu & $\rho$& 1.00 & 0.51 & 0.68 & 0.55 & 0.36 & 0.11 & 0.13 \\
(Rel.)      & $P$   & 1.00 & 0.41 & 0.60 & 0.45 & 0.45 & 0.08 & 0.10 \\
            & $v^x$ & 0.00 & 0.40 & 0.26 & 0.28 & 0.28 &-0.12 & 0.00 \\
            & $v^y$ & 0.00 &-0.15 &-0.46 &-0.58 &-0.07 &-0.07 & 0.00 \\
 & ${\cal{B}}^y/\sqrt{4\,\pi}$& 1.00 &-0.14 &-0.37 &-0.42 &-0.42 &-0.83 &-1.00 \\
\hline
Brio and Wu & $\rho$& 1.00 & 0.67 & 0.86 & 0.71 & 0.25 & 0.12 & 0.13 \\
(S92)       & $P (\times {10}^{-4})$   
                     & 1.00 & 0.45 & 0.73 & 0.50 & 0.50 & 0.09 & 0.10 \\
$\gamma=2.0$  & $v^x (\times {10}^{-3})$   
                    & 0.00 & 6.54 & 4.51  & 6.02  & 6.02  & -2.76 & 0.00 \\
            & $v^y (\times {10}^{-2})$   
                    & 0.00 &-0.24 &-1.15 &-1.58 &-1.58 &-0.20 & 0.00 \\
    &  ${\cal{B}}^y(\times {10}^{-2})/\sqrt{4\,\pi}$   
                    & 1.00 & 0.57 &-0.37 &-0.54 &-0.54 &-0.89 &-1.00 \\
\hline
Ryu and Jones& $\rho$& 1.00 & 0.68 & 0.74 & 0.91 & 0.61 & 0.32 & 0.40 \\
(RJ95)        & $P (\times {10}^{-4})$   
                     & 1.00 & 0.53 & 0.60 & 0.86 & 0.86 & 0.28 & 0.40 \\
$\gamma=5/3$ & $v^x (\times {10}^{-3})$   
                      & 0.00 & 6.66 & 4.62 & 3.03 & 3.03 &-5.61 & 0.00 \\
             & $v^y (\times {10}^{-2})$   
                       & 0.00 &-0.90 &-1.48 &-1.24 & -1.24 & -0.80 & 0.00 \\
    &  ${\cal{B}}^y(\times {10}^{-2})/\sqrt{4\,\pi}$   
                    & 1.00 & 0.28 &-0.34 &-0.24 &-0.24 &-0.44 &-1.00 \\
\hline
\end{tabular}
\end{table}

\newpage
\clearpage
\newpage
\clearpage

\newpage

\section{1D and 2D Test Cases - Kerr Spacetime}

To validate the general relativistic aspects of the code we require code tests
that involve magnetic fields and the Kerr geometry.  Test problems with analytic
solutions include one dimensional magnetized Bondi inflow, and a two-dimensional
Bondi flow along a split monopole field; in these tests the field does not
affect the dynamics of the problem.  We also perform a one-dimensional
magnetized inflow test problem where the field does act dynamically (Gammie
1999).  Finally, we have considered two-dimensional magnetized constant-$l$ tori
that are subject to the magnetorotational instability.  While there is no
analytic solution to this problem, the results can be compared to those obtained
in non-relativistic simulations.

\subsection{Magnetized Bondi Flow and Split Monopole Tests}

The analytic expressions for the hydrodynamic Bondi solution of HSWa are
unchanged in the presence of a radial magnetic field (see proof in Appendix A).
However, this result does not necessarily hold for a numerical solution to
the Bondi problem.  A non-trivial test of the magnetic components of the code
consists, therefore, in maintaining the equilibrium Bondi solution for {\it any}
magnitude of radial magnetic field.  The magnetized Bondi problem is set up in
the Schwarzschild metric ($a=0$).  The radial grid extends from just outside the
event horizon at $r=2.20\,M$ to $r=25.0\,M$; the critical point is at
$r_{crit}=8.0$.  Grids of $64$, $128$, $256$, $512$, and $1024$ points are used,
and the magnetic field at the critical point is set to $\beta=100,10$, and $1$,
where the $\beta$ parameter sets the grid-averaged ratio of gas to magnetic
pressure.  The Bondi solution is generated in the initialization routine by a
numerical root-finder on a logarithmically scaled grid of $1024$ points.  This
solution is then sampled at the required resolution for the particular test run
(to ensure proper alignment of the coarser grids).  To complete the test, the
unmagnetic case is also presented.  The solution is allowed to evolve to a time
sufficient to allow numerical convergence on each grid ($t_{final}=100\,M$).

\begin{figure}[ht]
     \epsscale{0.4}
     \plotone{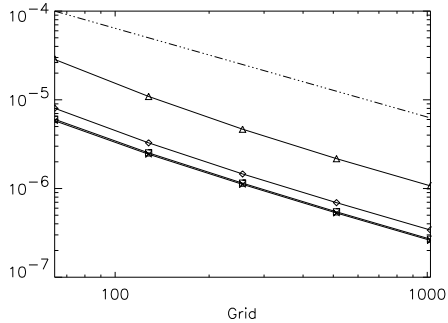}
     \caption{\label{Bondierror} 
     Plot of spatial error for Bondi test, for values of $\beta = 100$
     (crosses), $10$ (diamonds), $1$ (triangles) and the unmagnetized case 
     (squares) for variable $d$ .}
\end{figure}

Figure \ref{Bondierror} shows the results of spatial accuracy tests. 
The spatial accuracy of variable ${\cal X}$ is
obtained from the $L_1$ norm over an interval $r_{min} \le r \le r_{max}$,
\begin{equation}
L_1({\cal X}) = {1\over r_{max}-r_{min}}
 \int{\|{\cal X}_{final}-{\cal X}_{0}\| dr},
\end{equation}
where ${\cal X}_{final}$ denotes the converged solution, and ${\cal X}_{0}$ the
original solution, as described above.  Different regions of the flow are
dominated by numerical errors either from a particular routine in the numerical
solver or a particular code variable.  Here, the bounds of integration were
chosen to range from $r_{min}=7.0$ to $r_{max}=25.0$, which represents a region
of the flow where pressure error dominates.  Figure \ref{Bondierror} is for the
variables $d$, and is typical of error curves for all code variables.  The code
is clearly first-order accurate for these variables.  

In the split monopole test, we initialize a 2D $(r,\theta)$ grid with the radial
Bondi solution, but the radial magnetic field has opposite polarity above and
below the equator, hence establishing a current sheet along the equator.  This
test is made into a numerical stability test by applying small enthalpy
perturbations to the initial state (without this, the code is simply a set of 1D
radial tests).  The test is repeated with increasingly large values of magnetic
field, until the code breaks.  There is no known analytic solution for the
evolution of a current sheet, but it can be used as a probe of the code's
stability in the presence of strong, fluctuating current sheets.  The split
monopole test was carried out with $\beta=100,10,1$ at the critical point, and a
grid of $180^2$ scaled logarithmically in both the $r$ and $\theta$ coordinates,
with an initial $1 \%$ random enthalpy perturbation.  The test was allowed to
run to $t=1250M$ ($10$ free-fall times).  For $\beta =100, 10$, the code
variables show no visible change over the course of the run; the enthalpy
perturbations do not grow, nor do they trigger a numerical instability.  For
$\beta = 1$, the late-time solution shows evidence of reconnection at the
equator near the horizon, and erratic behavior. The effective $\beta$ where 
reconnection occurs is of order ${10}^{-1}$.  Given the moderate resolution,
the results are consistent with the spatial accuracy tests in that the region
near the horizon requires high resolution in order to minimize numerical error.

\newpage
\subsection{Magnetized Gammie Inflow}

Gammie (1999) describes a simplified 1D model for inflow from the inner
edge of an accretion disk in the Kerr geometry that serves as an
excellent test problem for relativistic MHD.  This solution is similar
to the Weber \& Davis (1967) rotating magnetized wind which was used as
a test problem Stone {\it{et al.}} (1992).  The model assumes that the
inflow is cold ($h \approx 1$), time-steady, purely radial, confined to
the equatorial plane, and initiated from the edge of an accretion disk
located at the marginally stable orbit ($r_{mso}$).  At $r_{mso}$ the
solution is assumed to satisfy the conditions $U^r(r_{mso})=0$ and
$\Omega_F \equiv V^\phi(r_{mso}) = 1/({r_{mso}}^{3/2}+a)$.  The flow is
characterized by a fast magnetosonic critical point.  The numerical
procedure to find a particular solution to this inflow is described in
Appendix D.

Our test consists in replicating the solution described by Gammie {\it{et al.}}
(2002) for a flow near a Kerr black hole with $a=0.5$ ($r_{horizon}=1.866$,
$r_{mso}=4.233$, $\Omega_F = 0.108588$), with ${\cal B}^r=-0.5$, $F_M=-1$, $F_L=
-2.815344$, $F_E=-0.9083782$.  The critical point for this test case is
$(r_{crit},U^r_{crit}) = (3.616655,-0.04054696)$.

\begin{figure}[ht]
     \epsscale{0.6}
     \plotone{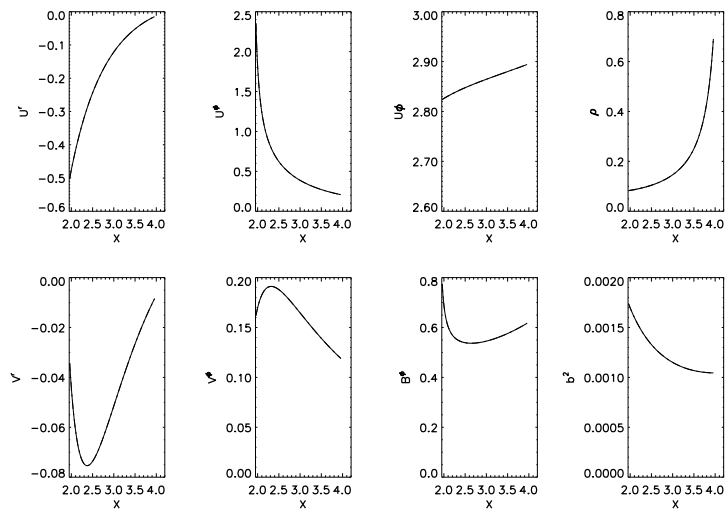}
     \caption{\label{G99M1A} 
     Plot of four-velocity components, density, and magnetic pressure 
     for the Magnetized Gammie Inflow test, with overlaid plots for
     $t = 0$ and $t = 15 M$. Grid 1024 points.}
\end{figure}

\begin{figure}[ht]
     \epsscale{0.4}
     \plotone{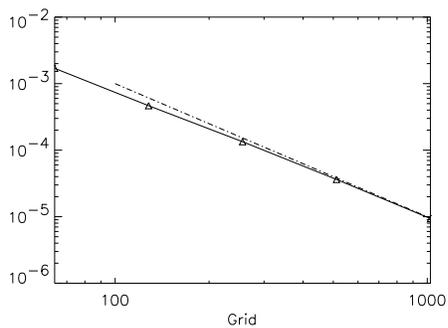}
     \caption{\label{G99ERR} 
     Plot of spatial error for Magnetized Gammie Inflow,
     (left) variable $d$ and
     (right) variable ${\cal B}^\phi$.}
\end{figure}

Figure \ref{G99M1A} shows a plot of the solution for this test problem on a
logarithmically scaled grid of $1024$ points.  The inner edge of the grid is at
$r_{min}=2.00$ and the outer edge of the grid is at $r_{max}=4.04$.  In
Boyer-Lindquist coordinates, the event horizon cannot be part of the
computational domain because of the coordinate singularity.  In the Gammie
inflow test, the marginally stable orbit, which is the formal outer boundary of
the problem, is also excluded since the density is divergent as $r$ approaches
$r_{mso}$ from below.  The plot overlays the late-time solution (after $1000$
time steps) on the initial state, but the two curves are indistinguishable at
the chosen plot scale.  Figure \ref{G99ERR} shows the results of convergence
tests carried out analogously to those for the Bondi problem, 
with the range of integration going from $r_{min}=2.7$ to $r_{max}=4.04$.  
Here we have an example of a highly-resolved near-horizon region, and the 
convergence rate is close to second order.

\newpage
\clearpage

\newpage
\subsection{Constant-$l$ Disks}

The final set of tests involve constant-$l$ disks ($l =-U_\phi/U_t$, the
specific angular momentum), solutions of the axisymmetric GR
hydrodynamic equations described in HSWa.  Here, we add weak poloidal magnetic
field loops that overlay the hydrodynamic solution; this will trigger the MRI.
We quantify the strength of the magnetic field through the $\beta$-parameter,
the ratio of the volume-averaged gas pressure to magnetic pressure.

The initial magnetic field is obtained from the definition
of $F_{\mu \nu}$ in terms of the $4$-vector potential, $A_\mu$,  
\begin{equation}\label{avec.1}
F_{\mu \nu} = \partial_\mu A_{\nu} - \partial_\nu A_{\mu} .
\end{equation}
Restricting the field to poloidal loops is done by using
$A_{\mu} = (A_t,0,0,A_\phi)$, where
\begin{equation}\label{vecpot}
A_\phi = 
\cases{
k (\rho-\rho_{min}) & for $\rho \ge \rho_{min}$ \cr
0 & for $\rho < \rho_{min}$}.
\end{equation}
Using (\ref{avec.1}), it follows that ${\cal{B}}^r = -\partial_\theta
A_{\phi}$ and ${\cal{B}}^\theta = \partial_r A_{\phi}$.  This choice of
$A_{\phi}$ produces poloidal field loops that coincide with isodensity
contours.  The constant $k$ is set by the input parameter $\beta$.  The
constant $\rho_{min}$ sets a suitable minimum density within the disk,
and it is chosen to keep the initial magnetic field away from the
outer edge of the disk.

\begin{figure}[hb]
     \epsscale{1.0}
     \plotone{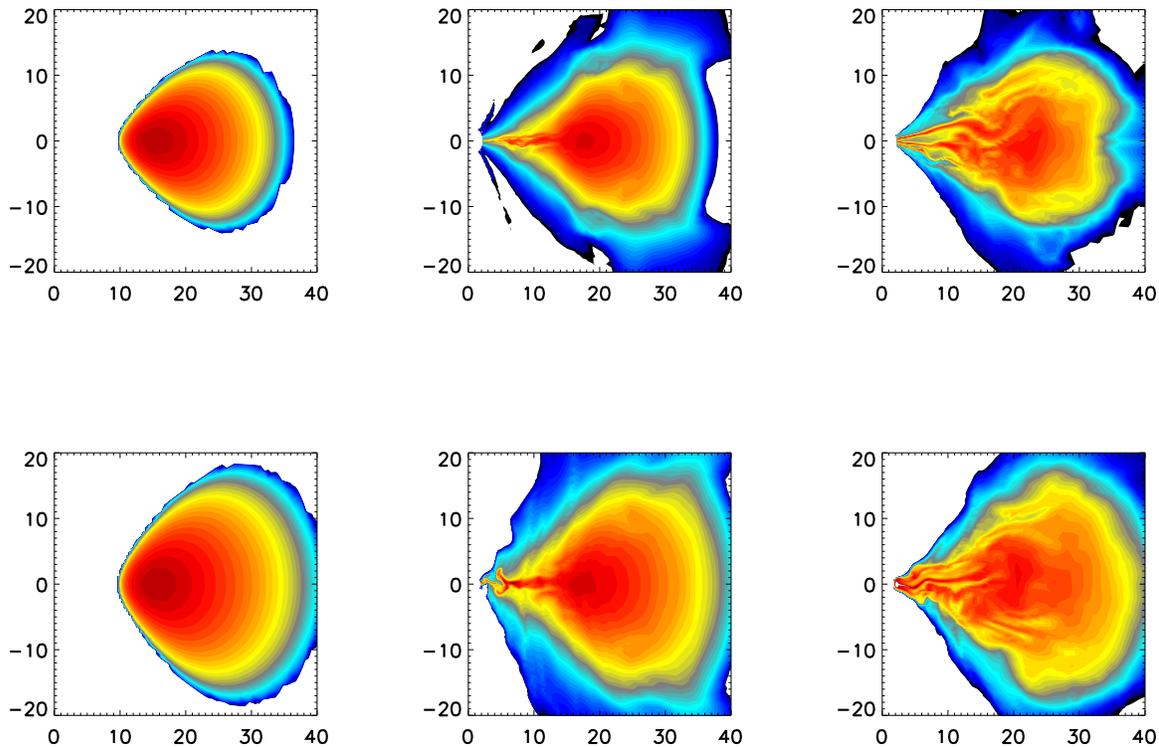}
     \caption{\label{MRI} 
     Plot of log density for the Schwarzschild (top row) 
     and Kerr (bottom row) constant-$l$ disks.  The left column is the
     initial state, the middle column is from $t=1.5$ orbits, and the 
     right column corresponds to $t=3.0$ orbits at the initial pressure 
     maximum.}
\end{figure}

An axisymmetric (2D) magnetized constant-$l$ torus was set up in both
the Schwarzschild and Kerr metrics.  For the Schwarzschild case, the
disk has a specific angular momentum $l=4.5$, an initial pressure
maximum at $r = 15.3 M$, and an orbital period at the pressure maximum
of $T_{orb} =376 M$.  For the Kerr case, the black hole is maximally
rotating, $a=1$, and the disk has a prograde specific angular momentum
$l=4.3$, an initial pressure maximum at $r = 15.4 M$, and an orbital
period at the pressure maximum of $T_{orb} =386 M$.  This particular
choice of parameters yields a disk that is similar to the Schwarzschild
case, although the equipotentials which define the overall disk
structure are notably different, as can be seen in the left column of
Figure \ref{MRI}.  For both tests the average field strength is $\beta
= 100$, the initial state is perturbed with random $1 \%$ enthalpy
fluctuations.  The simulations are run for $10$ orbits.  The MRI
develops after a few orbits (center column, Fig. \ref{MRI}), and is
soon fully developed (right column, Fig. \ref{MRI}).  By the tenth
orbit, the MRI-induced turbulence has settled down considerably, as is
to be expected since the imposition of axisymmetry precludes the
development of the azimuthal modes which sustain the MRI.
The observed qualitative differences between the Schwarzschild and
Kerr cases can be attributed to the different shapes of the potential cusp
near the horizon.

The density plots bear a strong resemblance to the previous
axisymmetric MRI studies of Hawley (2000) in a pseudo-Newtonian
potential; this indicates that the numerical solver is well behaved in
the moderately strong gravitational fields that exist near the pressure
maximum, where the implied code comparison is taking place.  Since MRI
studies are the main area of application of this code, it is reassuring
that we can both trigger the MRI in a general relativistic treatment of
MHD, and that we qualitatively reproduce established results.  Gammie
{\it et al.} (2002) obtain similar results with their MRI torus test.

\newpage
\section{Discussion}

In this paper we document the development of a General Relativistic MHD
code that is a direct extension of the general relativistic
hydrodynamic solver of Hawley, Smarr and Wilson (1984b).  As before,
the equations of relativistic MHD are written in a nonconservative
form, supplemented here by the relativistic induction equation.  The
magnetic field is added to the definition of the inertia for the four
momentum and the magnetic field adds several source terms to its
evolution.

The equations are evolved on a staggered grid using the time-explicit
model of ``transport $+$ source.'' The induction equation is evolved
using the Constrained Transport structure of EH88.  We implement a
version of CT that is inspired by the Method of Characteristics
technique employed in the popular ZEUS code (Stone \& Norman 1992), yet
is simpler to implement in a general relativistic framework.

We have performed a number of 1D and 2D tests to validate the code and
to note its limitations.  The tests include a suite of non-relativistic
and relativistic MHD shock tube tests, the fast and slow rarefaction
and slow magnetosonic shock tests of Komissarov (1999),  and simple
Alfv\'en pulse propagation.  The formulation for these test and their
results are documented to establish a basis for comparison against
other existing and future codes.

The main shortcoming of our code revealed by these tests is due to the
already-known limitations of the artificial viscosity algorithm of the
underlying hydrodynamic solver.  Improvements to this algorithm are an
ongoing area of research, but is important to stress that these
limitations do not pose an undue restriction on our main area of
study:  dynamical, MHD-driven accretion flows around spinning black
holes.

Although the set of general relativistic black hole accretion test
problems is limited, we have also verified that the code reproduces the
standard results for Bondi flow in the presence of magnetic fields, and
have observed the triggering and evolution of the MRI in thick
accretion tori for both the Kerr and Schwarzschild black holes.  The
MRI evolution can be compared qualitatively with pseudo-Newtonian
simulations (e.g.  Hawley 2000).  One new test problem with an analytic
solution is the magnetized Gammie (1999) inflow test.  This solution
examines the behavior of a dynamically important magnetic field in the
strong-field region of the Kerr metric.  The results of these tests
provide evidence that this new solver is correctly reproducing
magnetized flows outside the event horizon of a Kerr black hole.  This
new GR MHD code should provide a valuable tool with which to study
accretion flows in the Kerr metric.

\acknowledgements{This work was supported by NSF grant AST-0070979 and
NASA grant NAG5-9266.  We thank Charles Gammie with whom we collaborated
to develop and apply a suite of code tests for general relativistic MHD.}

\newpage
\appendix

\section{Proof or Various Results}

Some of the following proofs use two identities for contractions of the 
Levi-Civita
tensor,
\begin{equation}
\label{levi.1}
\epsilon^{\alpha\,\beta\,\delta\,\gamma}\epsilon_{\alpha\,\lambda\,\mu\,\nu}=
-{\delta^{\beta\,\delta\,\gamma}}_{\lambda\,\mu\,\nu},
\end{equation}
\begin{equation}
\label{levi.2}
\epsilon^{\alpha\,\beta\,\delta\,\gamma}\epsilon_{\alpha\,\beta\,\mu\,\nu}=
-2\,{\delta^{\delta\,\gamma}}_{\mu\,\nu},
\end{equation}
and identities for the $\delta$-symbol,
\begin{equation}\label{delta2}
{\delta^{\alpha\,\beta}}_{\sigma\,\rho}
=\delta^{\alpha}_{\sigma}\delta^{\beta}_{\rho}-
\delta^{\alpha}_{\rho}\delta^{\beta}_{\sigma},
\end{equation}
\begin{equation}\label{delta3}
{\delta^{\alpha\,\beta\,\delta}}_{\lambda\,\mu\,\nu}
=\delta^{\alpha}_{\lambda}\delta^{\beta}_{\mu}\delta^{\delta}_{\nu}-
\delta^{\alpha}_{\mu}\delta^{\beta}_{\lambda}\delta^{\delta}_{\nu}+
\delta^{\alpha}_{\mu}\delta^{\beta}_{\nu}\delta^{\delta}_{\lambda}-
\delta^{\alpha}_{\nu}\delta^{\beta}_{\mu}\delta^{\delta}_{\lambda}+
\delta^{\alpha}_{\nu}\delta^{\beta}_{\lambda}\delta^{\delta}_{\mu}-
\delta^{\alpha}_{\lambda}\delta^{\beta}_{\nu}\delta^{\delta}_{\mu}.
\end{equation}

To prove (\ref{fmunudef.2}), it is easiest to work backwards from the
result. Expand the right-hand side of (\ref{fmunudef.2}) using (\ref{bdef.2})
and (\ref{dual.2}); the resulting expression simplifies through identities 
(\ref{levi.1}) and (\ref{delta3}), followed by the velocity 
normalization condition,
infinite conductivity, and antisymmetry of $F^{\mu \nu}$:
\begin{eqnarray}
\nonumber
\epsilon_{\alpha\,\beta\,\mu\,\nu}\,B^{\alpha}\,U^\beta
& = & {1 \over 2}\,
\epsilon^{\alpha\,\delta\,\epsilon\,\gamma}
 \epsilon_{\alpha\,\beta\,\mu\,\nu}\,U_\delta\,U^\beta\,
 F_{\epsilon\,\delta} \\
\nonumber & = &
-{1 \over 2} {\delta^{\delta\,\epsilon\,\gamma}}_{\beta\,\mu\,\nu}\,
U_\delta\,U^\beta\,F_{\epsilon\,\delta} \\
\nonumber & = &
-{1 \over 2}\left( U_\beta\,U^\beta\,F_{\mu\,\nu}-U_\mu\,U^\beta\,F_{\beta \nu}+
U_\mu\,U^\beta\,F_{\nu\,\beta}-U_\nu\,U^\beta\,F_{\mu\,\beta}+
U_\nu\,U^\beta\,F_{\beta\,\nu}-U_\beta\,U^\beta\,F_{\nu\,\mu}\right)\\
\nonumber & = &
F_{\mu\,\nu}.
\end{eqnarray}

To prove (\ref{tmunuem}), expand the EM component of tensor (\ref{tmndef}) 
using 
(\ref{fmunudef.2}), and simplify using identities (\ref{levi.1})-(\ref{delta3}),
followed by the velocity normalization condition,
and orthogonality condition (\ref{bnorm}):
\begin{eqnarray}
\nonumber
{T}^{\mu\,\nu}_{(EM)} & = & {1 \over 4\,\pi} 
\left(g^{\mu \gamma}\,F_{\gamma \alpha}\,F^{\nu \alpha} 
 - {1 \over 4} F_{\alpha \beta}\,F^{\alpha \beta}\,g^{\mu \nu} \right)\\
\nonumber
 & = & {1 \over 4\,\pi} 
\left[g^{\mu \gamma}\,
 \left(\epsilon_{\delta\,\epsilon\,\gamma\,\alpha}\,B^{\delta}\,U^\epsilon\right)\,
 \left(\epsilon^{\rho\,\sigma\,\nu\,\alpha}\,B_{\rho}\,U_\sigma\right) - {1 \over 4} 
 \left(\epsilon_{\delta\,\epsilon\,\alpha\,\beta}\,B^{\delta}\,U^\epsilon\right)\,
 \left(\epsilon^{\rho\,\sigma\,\alpha\,\beta}\,B_{\rho}\,U_\sigma\right)\,g^{\mu \nu} \right]\\
\nonumber
 & = & -{1 \over 4\,\pi} 
\left(g^{\mu \gamma}\,
\delta^{\rho\,\sigma\,\nu}_{\delta\,\epsilon\,\gamma}\,B^{\delta}\,U^\epsilon\,B_{\rho}\,U_\sigma-
{g^{\mu \nu} \over 2}\,\delta^{\rho\,\sigma}_{\delta\,\epsilon}\,B^{\delta}\,U^\epsilon
\,B_{\rho}\,U_\sigma\right)\\
\nonumber
 & = & -{1 \over 4\,\pi} 
\left(-{g^{\mu\,\nu}\,B^{\alpha}\,B_{\alpha} \over 2}+
B^{\mu}\,B^{\nu}-B^{\alpha}\,B_{\alpha}\,U^\mu\,U^\nu\right)\\
\nonumber
 & = & 
{1 \over 2}\,g^{\mu\,\nu}\,{\|b\|}^2+U^\mu\,U^\nu\,{\|b\|}^2
-b^{\mu}\,b^{\nu}.
\end{eqnarray}

To prove (\ref{max2.b}), the second of Maxwell's equations,
${\nabla}_{\mu}{}^*F^{\mu\,\nu} =0$, can be 
rewritten by substituting definitions, and
using identities (\ref{levi.2}) and (\ref{delta2}):
\begin{eqnarray}
\nonumber
{\nabla}_{\alpha}{}^*F^{\alpha\,\beta} & = & {1 \over 2}
{\nabla}_{\alpha}\,\epsilon^{\alpha\,\beta\,\delta\,\gamma}\,
F_{\delta\,\gamma}\\
\nonumber
 & = & {1 \over 2}
{\nabla}_{\alpha}\,\epsilon^{\alpha\,\beta\,\delta\,\gamma}\,
\epsilon_{\rho\,\sigma\,\delta\,\gamma}\,b^{\rho}\,U^{\sigma}\\
\nonumber
 & = & -{\nabla}_{\alpha}\,\delta^{\alpha\,\beta}_{\rho\,\sigma}\,b^{\rho}\,
 U^{\sigma} ,
\end{eqnarray}
which simplifies, using (\ref{delta2}), to 
${\nabla}_{\alpha}\left(U^\alpha\,b^\beta- b^\alpha\,U^\beta\right) =0$.

To prove (\ref{bsqdef}), we begin with an identity derived from 
the expression for $W$ (\ref{gred}),
\begin{equation}
 W^2 \,\alpha^{-2} = -g^{tt}\,W^2 = -{(g_{tt} + 2\,g_{t \phi}\,V^\phi +
 g_{r r}\,{(V^r)}^2 + g_{\theta \theta}\,{(V^\theta)}^2 + 
 g_{\phi \phi}\,{(V^\phi)}^2)}^{-1} ,
\end{equation}
or
\begin{equation}
 g_{tt} + g_{r r}\,{(V^r)}^2 + g_{\theta \theta}\,{(V^\theta)}^2 + 
 g_{\phi \phi}\,{(V^\phi)}^2 = -{\alpha^2 \over W^2 } - 2\,g_{t \phi}\,V^\phi .
\end{equation}
Now expand ${\|b\|}^2$,
\begin{equation}
{\|b\|}^2 = g_{tt} {(b^t)}^2+ g_{r r}\,{(b^r)}^2 + g_{\theta \theta}\,{(b^\theta)}^2 + 
 g_{\phi \phi}\,{(b^\phi)}^2 + 2\,g_{t \phi}\,b^t\,b^\phi ,
\end{equation}
and substitute
\begin{equation}
b^i = { {\cal{B}}^i \over \sqrt{4 \pi}\,W\,\sqrt{\gamma}}+b^{t}\,V^i
\end{equation}
for the spatial components of the magnetic field. Making use of the above
identity, it follows that
\begin{eqnarray}
{\|b\|}^2 & = & 
-\left[{\alpha^2 \over W^2 } + 2\,g_{t \phi}\,V^\phi\right]\,{(b^t)}^2+
2\,g_{t \phi}\left[
{{\cal{B}}^\phi \over \sqrt{4 \pi}\,W\,\sqrt{\gamma}}+b^{t}\,V^\phi \right]\,b^t\\
\nonumber & + &
{2 \over \sqrt{4 \pi}\,W\,\sqrt{\gamma}}\,\left[
  g_{rr}\,{\cal{B}}^r\,V^r +g_{\theta \theta}\,{\cal{B}}^\theta\,V^\theta+
  g_{\phi \phi}\,{\cal{B}}^\phi\,V^\phi \right]\,b^t\\
\nonumber & + &
{1 \over 4 \pi\,W^2\,\gamma}
\left[g_{rr}\,{({\cal{B}}^r)}^2 +g_{\theta \theta}\,{({\cal{B}}^\theta)}^2+
  g_{\phi \phi}\,{({\cal{B}}^\phi)}^2 \right] .
\end{eqnarray}
Now substitute the expression for $b^t$ (\ref{bdef.5}) and simplify,
noting that 
\begin{equation}
g_{\phi \phi}\,V^\phi + g_{t \phi} = {g^{tt}\,V^\phi - g^{t \phi} \over 
   g^{tt}\,g^{\phi \phi} - {(g^{t \phi})}^2}
\end{equation}
to get, after a few steps, expression (\ref{bsqdef}).

To derive (\ref{alfvel}) we first obtain the components of the
fluid-frame magnetic field, which are, to leading order,
\begin{eqnarray}
\nonumber
b^t & \approx & g_{xx}\,{W\,V^x\,{\cal{B}}^x \over \alpha^2\,\sqrt{4\,\pi\,\gamma}} ,\\
\nonumber
b^x & \approx & {W\,{\cal{B}}^x \over \sqrt{4\,\pi\,\gamma}} ,\\
\nonumber
b^y & \approx & {{\cal{B}}^y \over \sqrt{4\,\pi\,\gamma}\,W}+
g_{xx}\,{W\,V^x\,V^y\,{\cal{B}}^x \over \alpha^2\,\sqrt{4\,\pi\,\gamma}} ,\\
\nonumber
{\|b\|}^2 & \approx & g_{xx}\,
{\left({{\cal{B}}^x \over \sqrt{4\,\pi\,\gamma}}\right)}^2 ,
\end{eqnarray}
and also note the approximate value for $W$,
\begin{equation}
W^{-2}  \approx 1- {g_{xx} (V^x)}^2 \, \alpha^{-2} .
\end{equation}
We then expand the momentum
equation and substitute the expressions for $b_y$ and $b^t$.
The left-hand side of the momentum equation reads
\begin{eqnarray}
\nonumber
\partial_t\,\left(S_y - \alpha\,b_y\,b^t\right) & = & 
\partial_t\,\left(g_{yy}\,(\rho\,h+{\|b\|}^2)\,W^2\,\alpha^{-1}\,V^y - 
g_{yy}\,\alpha\,b_y\,b^t\right) \\
\nonumber
 & \approx & 
\partial_t\,\left(
g_{yy}\,\rho\,h\,{W^2 \over \alpha}\,V^y + 
g_{yy}\,{\|b\|}^2\,{W^2 \over \alpha}\,V^y\,
 \left[1-{g_{xx}\,{(V^x)}^2 \over \alpha^2}\right] - 
{g_{xx}\,g_{yy}\,{\cal{B}}^x\,{\cal{B}}^y\,V^x \over
 \alpha\,{\left(\sqrt{4\,\pi\,\gamma}\right)}^2}
\right) \\
\nonumber
 & \approx & 
{g_{yy}\,(\rho\,h\,W^2+{\|b\|}^2) \over \alpha }\,\partial_t\,\left(V^y\right) - 
{g_{xx}\,g_{yy}\,{\cal{B}}^x\,V^x \over \alpha\,{\left(\sqrt{4\,\pi\,\gamma}\right)}^2}\,
 \partial_t\,\left({\cal{B}}^y\right) .
\end{eqnarray}
Recall that $x$ stands for either $r$ or $\theta$, and 
$y$ stands for a transverse direction, either $r$, $\theta$, or $\phi$. 

As for the left hand side, we neglect the 
purely geometric terms and retain only
\begin{eqnarray}
\nonumber
-{1 \over \sqrt{\gamma}}\,\partial_x\,\sqrt{\gamma}\,
 \left(S_y\,V^x - \alpha\,b_y\,b^x\right) 
& = & 
-{1 \over \sqrt{\gamma}}\, \partial_x\,\sqrt{\gamma}\,
\left(
 g_{yy}\,(\rho\,h+{\|b\|}^2)\,{W^2 \over \alpha}\,V^y\,V^x - 
 \alpha\,g_{yy}\,b^y\,b^x\right) \\
\nonumber
& \approx & -{1 \over \sqrt{\gamma}}\, \partial_x\,\sqrt{\gamma}\,
\left(
 g_{yy}\,\rho\,h\,{W^2 \over \alpha}\,V^y\,V^x - 
 \alpha\,g_{yy}\,{{\cal{B}}^x\,{\cal{B}}^y \over
 {\left(\sqrt{4\,\pi\,\gamma}\right)}^2}\right) \\
\nonumber
& \approx & -{g_{yy}\,\rho\,h\,W^2\,V^x \over \alpha}\partial_x\,
\left(V^y\right) + {\alpha\,g_{yy}\,{\cal{B}}^x \over
 {\left(\sqrt{4\,\pi\,\gamma}\right)}^2}\,\partial_x\,
\left({\cal{B}}^y\right)  .
\end{eqnarray}

Combine these results and replace $\partial_t\,\left({\cal{B}}^y\right)$
with the results for the induction
equation,
\begin{eqnarray}
\nonumber
\partial_t\,\left(V^y\right) +
{\rho\,h\,W^2\,V^x \over (\rho\,h\,W^2+{\|b\|}^2)}\partial_x\,\left(V^y\right)  
& \approx &
{g_{xx}\,{\cal{B}}^x\,V^x \over 4\,\pi\,\gamma(\rho\,h\,W^2+{\|b\|}^2) }\,
 \partial_t\,\left({\cal{B}}^y\right) \\
\nonumber &  +&  {\alpha^2\,{\cal{B}}^x \over 
 4\,\pi\,\gamma(\rho\,h\,W^2+{\|b\|}^2)}\,\partial_x\,
\left({\cal{B}}^y\right) \\
\nonumber
& \approx &
{g_{xx}\,{\cal{B}}^x\,V^x \over 4\,\pi\,\gamma(\rho\,h\,W^2+{\|b\|}^2) }\,
\left({\cal{B}}^x\,\partial_x\,\left(V^y\right)- 
V^x\,\,\partial_x\left({\cal{B}}^y \right)\right)\\
\nonumber & + &
{\alpha^2\,{\cal{B}}^x \over 
 4\,\pi\,\gamma(\rho\,h\,W^2+{\|b\|}^2)}\,\partial_x\,
\left({\cal{B}}^y\right) .
\end{eqnarray}
By regrouping these terms and substituting the approximate expressions for 
$W$ and ${\|b\|}^2$
given above, equation (\ref{alfvel}) follows immediately. A similar
procedure yields expression (\ref{alfvel_phi}) for transverse velocities  
when $x \equiv \phi$.

To prove that the Bondi solution is unchanged by the presence of a radial
magnetic field, we revisit the calculations presented in HSWa. 
We are looking for a time-independent purely radial solution to the
equations of GRMHD in the presence of a constant CT radial magnetic field
${\cal{B}}^r = F_{\theta \phi} = B$
in the Schwarzschild metric. We will work with the more 
general formalism of equations (\ref{barcons}) and (\ref{tmncons}).

The equation of continuity (\ref{barcons}) is not dependent on magnetic fields, 
so
\begin{equation}
\label{bondi.1}
\nabla_\mu \left(\rho\,U^\mu\right) = 0 \rightarrow 
\nabla_r \left(\rho\,U^r\right) = 0,
\end{equation} 
leads to the same conserved quantity as in the hydrodynamic case,
\begin{equation}\label{bondi.2}
\sqrt{-g}\,\rho\,U^r = C_1.
\end{equation} 
The energy-flux conservation equation is the $t$-component of (\ref{tmncons}),
\begin{equation}\label{bondi.3}
\nabla_\mu {T^\mu}_t = 0 \rightarrow
\nabla_r {T^r}_t = 0 ,
\end{equation} 
and contains an explicit dependence on the magnetic field. However, by expanding
the $r-t$ component of the energy momentum tensor,
\begin{equation}\label{bondi.4}
{T^r}_t = \left(\rho\,h+{\|b\|}^2 \right)\,U^r\,U_t-b^r\,b_t ,
\end{equation} 
and using equations (\ref{fmunudef.3}) and (\ref{bdef.5}) to obtain the 
fluid-frame magnetic field,
\begin{equation}\label{bondi.5}
b^r = {{\cal B}^r \over \sqrt{4\,\pi}} {W \over \sqrt{\gamma}}\,,\,
b^t = -g_{rr}\,b^r\,V^r\,,\,
{\|b\|}^2 = g_{rr}\,{{(b^r)}^2 \over W^2} ,
\end{equation} 
we obtain after simple substitutions
\begin{equation}\label{bondi.4a}
{T^r}_t = \left(\rho\,h+{\|b\|}^2 \right)\,U^r\,U_t-{\|b\|}^2\,U^r\,U_t .
\end{equation} 
Clearly, the magnetic contributions cancel (this is simply restating the
general result that there is no net force due to magnetic fields parallel 
to the direction of motion).
We therefore recover the hydrodynamic expression,
\begin{equation}\label{bondi.6}
\nabla_r {T^r}_t = \left(\rho\,h\right)\,U^r\,U_t = 0,
\end{equation}
which leads to the same conserved quantities as in the hydrodynamic case,
\begin{equation}\label{bondi.7}
\sqrt{-g}\,\rho\,U^r\,h\,U_t = C_2,
\end{equation}
and
\begin{equation}\label{bondi.8}
h\,U_t = C_3.
\end{equation} 

\section{Amplitude of Alfv\'en Pulses}

To prove (\ref{alfgam}), we used the method of characteristics. We
write the momentum and induction equations as a matrix equation,
\begin{equation}
A\,\partial_x\,U + B\,\partial_t\,U = 0
\end{equation}
where $U(x,t)={\left(V^y(x,t),{\cal B}^y(x,t)\right)}^T$ and
\begin{equation}
A = \left [\matrix{
{V^x \over (1+\eta^2)} & -{\eta^2 \over {\cal B}^x\, (1+\eta^2)} \cr
-{\cal B}^x & V^x \cr
}\right ] \, ,\,
B = \left [\matrix{
 1 & \chi   \cr
 0 & 1   \cr
 }\right ]\, ,\,\chi = -{\eta^2 V^x\over {\cal B}^x\, (1+\eta^2)} .
\end{equation}
subject to the initial state $U_0(x,0)={\left(f(x),0\right)}^T$
We obtain the eigenvalues from the roots of $F(\lambda)=\det{A-\lambda\,B}$,
which are simply the expressions for ${v_A}^{(\pm)}$, (\ref{alfven.5}).

To solve this system, we seek a matrix $P$ such that $P\,B\,U=z$ where
$z_i(x,t)=g_1(x-\lambda_i\,t)$ and $g(x,0)=P\,B\,U_0(x,0)$. The rows
of matrix $P$ satisfy $A^T-\lambda_i\,B^T\,P_i=0$. It is straightforward
to show that the matrix $P$ has the form,
\begin{equation}
P = \left [\matrix{
 \zeta & 1  \cr
-\zeta & 1  \cr
 }\right ]
\, , \,
\zeta =  {{\cal B}^x (1+\eta^2)\over \eta\,\sqrt{\eta^2 +W^{-2}}} .
\end{equation}
We obtain the results
\begin{eqnarray}
 \zeta\,V^y +  (1+\zeta\,\chi)\,{\cal B}^y & =& \zeta\,f(x-{v_A}^{-}\,t),\\
-\zeta\,V^y +  (1-\zeta\,\chi)\,{\cal B}^y & =&-\zeta\,f(x-{v_A}^{+}\,t),
\end{eqnarray}
which yield
\begin{eqnarray}
V^y(x,t)& =& \left({1-\zeta\,\chi\over 2}\right)\,f(x-{v_A}^{-}\,t)+
         \left({1+\zeta\,\chi\over 2}\right)\,f(x-{v_A}^{+}\,t),\\
{\cal B}^y(x,t)& =& {\zeta \over 2}\,\left(f(x-{v_A}^{-}\,t)-
          f(x-{v_A}^{+}\,t)\right).
\end{eqnarray}
As discussed in the text, the velocity amplitudes are clearly
asymmetrical, except for the case where $V^x=0$, and the magnetic
field amplitudes are always equal, but opposite in sign. It is
easy to show that the ratio
of the velocity amplitudes is indeed the ratio
of the gamma factors, $W({v_A}^{(\pm)})$. It is also possible to
verify by direct computation that the amplitudes of the pulses given
in the text agree with the above formulas.

\section{MHD Shocks}

Relativistic MHD shocks have been discussed by 
Anile (1989) and Komissarov (1999).  Their work builds upon the 
monograph of Lichnerowicz (1967) who systematically developed the relations for
GR MHD shock invariants.  We summarize here the relevant results from
Lichnerowicz, translating symbols to our own notation. We use these
results to construct our shock tests.

MHD shocks are discontinuities in one or more of the MHD variables whose
propagation through spacetime lies on a hypersurface $\Sigma$. It is useful in
the analysis of these shocks to construct a $4$-vector $n^\mu$ normal to
$\Sigma$.  Shocks can be characterized by invariant scalar and $4$-vector
quantities---the jump conditions---which are derived readily from the equations of
GRMHD.  MHD shocks can be assigned to one of two categories, tangential and
non-tangential, and this classification hinges on whether the invariants have
non-zero projections onto the normal $n^\mu$.  Slow and fast magnetosonic shocks
are examples of non-tangential shocks, so it is this type of shock that is of
interest here.

The analysis of non-tangential shocks was shown by Lichnerowicz to reduce to the
study of five scalar quantities, $\rho$, $h$, $\eta = b^\mu\,n_\mu$, ${\cal{A}}
= \rho\,U^\mu\,n_\mu$, and $\|b^2\|$ which are related through five 
positive-definite scalar invariants (derivable from the jump conditions),
\begin{eqnarray}
\label{shkinvarA}
\left[{\cal{A}}\right] & = & 0 ,\\
\label{shkinvarB}
\left[{\cal{B}}\right] & = & \left[h \, \eta \right] = 0,\\
\label{shkinvarH}
\left[{\cal{H}}\right] & = & \left[{\eta^2 \over {\cal{A}}^2}-
  {\|b^2\| \over \rho^2}\right] = 0,\\
\label{shkinvarL}
\left[{\cal{L}}\right] & = & \left[\alpha + {q \over {\cal{A}}^2}\right]  = 0,\\
\label{shkinvarK}
\left[{\cal{K}}\right] & = & \left[
  \left(\rho^2+{\cal{A}}^2\right)  {h^2 \over \rho^2}+
  k^2 \left( 2\,{h \over \rho} - {\cal{H}} \right)\right] = 0 ,
\end{eqnarray}
where $\left[{\cal{X}}\right] = {\cal{X}}_{+} - {\cal{X}}_{-}$ 
indicates the change in a quantity ${\cal{X}}$ across the shock, and
\begin{eqnarray}
q & = & P + {\|b^2\| \over 2},\\
k^2 & = & \|b^2\| - {\cal{A}}^2\,{\cal{H}},\\
\alpha & = & {h \over \rho} - {\cal{H}}.
\end{eqnarray}

Magnetosonic shocks are characterized by the parameter $\alpha$.  The
determination as to whether a magnetosonic wave is fast or slow hinges on the
sign of the pre- and post-shock values of parameter $\alpha$.  Physically
realizable shocks are characterized by $\alpha_{(+)} \, \alpha_{(-)} > 0$.  A
slow shock has $\alpha_{(+)} < \alpha_{(-)} < 0$ and a fast shock has $0
<\alpha_{(+)}< \alpha_{(-)}$.

To construct an initial state, we solve the scalar invariants using the
Mathematica FindRoot function.  First, we postulate a left state in the rest
frame of the shock, where the normal vector $n^\mu$ has an especially simple
expression ($n^\mu=(0,1,0,0)$).  Next, we set up the scalar invariants as a
system to be solved by the FindRoot function, supplying an initial guess for the
right state.  This is an iterative procedure, requiring adjustments to the
initial guess.  Once a solution is found, it is possible to construct a moving
shock by repeating the above process with a small non-zero shock speed, for
which the normal vector now has the form $n^\mu=W(v_{sh})\,(v_{sh},1,0,0)$.
Once a solution is found the process is repeated, gradually increasing the shock
speed and using the previous converged right state as a starting point for the
solution at the new shock speed.  This is a pragmatic solution to a problem that
has a high degree of complexity, and it allows us to set up a family of closely
related shock solutions to aid in testing.

Given that the FindRoot function has converged on a right state, it is possible
to compute the parameter $\alpha$, and classify the shock.  It is also prudent
to verify that the left and right states are true solutions by verifying that
they satisfy the shock invariants by direct computation.  Once this is done, the
left and right states can be transferred directly to a numerical grid.

\section{Determining Conserved Quantities for Magnetized Gammie Inflow}

The magnetized Gammie inflow is characterized by three conserved quantities,
\begin{eqnarray}
F_M & = & 2\,\pi\,r^2\,U^r\,\rho ,\\
F_L & = & 2\,\pi\,r^2\,{T^r}_{\phi} = 2\,\pi\,r^2\,\left[
 \rho\,U^r\,U_\phi-{g^{rr}\,g^{\theta \theta} \over 4\,\pi}\,
 F_{r \theta}\,F_{\theta \phi} \right],\\
F_E & = &-2\,\pi\,r^2\,{T^r}_{t} = -2\,\pi\,r^2\,\left[
\rho\,U^r\,U_t+{g^{rr}\,g^{\theta \theta} \over 4\,\pi}\,
 F_{r \theta}\,F_{t \theta} \right],
\end{eqnarray}
which are readily derived from the equations of baryon and
momentum conservation. Note that Gammie writes 
${\cal{D}} \equiv r^2\,g^{rr}\,g^{\theta \theta}$.
From the induction equation, 
\begin{eqnarray}
\partial_r {\cal{B}}^r & = & 0 ,\\
\nonumber
\partial_r \left(V^\phi\,{\cal{B}}^r-V^r\,{\cal{B}}^\phi\right) & = & 0,
\end{eqnarray}
we obtain ${\cal{B}}^r=-F_{\theta \phi}=const$, and $F_{t \theta} =
V^\phi\,{\cal{B}}^r-V^r\,{\cal{B}}^\phi= \Omega_F\,F_{\theta \phi}$
which follows from the boundary
condition. Infinite conductivity ($U^\mu\,F_{\mu \nu}=0$)
yields
\begin{equation}
F_{r \theta} = {F_{\theta \phi} \over U^r} \left(U^\phi-\Omega_F\,U^t \right).
\end{equation}
Finally, the velocity normalization condition yields
\begin{equation}
-1 = g^{tt}\,{(U_t)}^2+2\,g^{t \phi}\,U_t\,U_\phi+g^{\phi\phi}\,{(U_\phi)}^2+
{(U^r)}^2/g^{rr}
\end{equation}
where the mix of upper and lower indices was made to 
coincide with code variables.

The critical point is a saddle point in the two-dimensional parameter space of
the energy function $F_E \equiv F_E(r,U^r;F_L)$.  It is apparent that $F_E$ as
given above is not a simple function of these parameters, and it is necessary to
substitute the other expressions in order to obtain the desired form for $F_E$.
Since the resulting expression for $F_E$ is quite lengthy, we will only sketch
the procedure.  The process of substitution and elimination described here was
carried out using Mathematica.

First, the conserved quantity $F_M$ is used to obtain 
$\rho(r,U^r) = F_M/(2\,\pi\,r^2\,U^r)$.
Next, the conserved quantity $F_L$ is used to isolate $U_\phi$,
\begin{equation}
U_\phi = \left[{F_L \over 2\,\pi\,r^2\,U^r}+{{\cal{D}} \over 4\,\pi\,r^2}\,
F_{\theta \phi}\,F_{r \theta}\right]\,{1 \over \rho U^r}.
\end{equation}
Now substitute the expressions for $\rho(r,U^r)$ and $F_{r \theta}$ 
expressed in terms of $U_{\phi}$ and $U_t$, to yield an expression of the form
$U_\phi = A + B\,U_t$ where $A$ and $B$ are expressions involving the constants
$F_L$, $F_M$, $F_{\theta \phi}$, and $\Omega_F$, as well as $U^r$ and metric
terms.  Now substitute this expression for $U_\phi$ in the velocity
normalization condition and ``solve'' the resulting quadratic for $U_t$, denote
this solution $(U_t)^{\pm}$ to emphasize that there are two distinct roots.
Finally, take the conserved quantity $F_E$ and substitute the expression for
$F_{r \theta}$ (expressed in terms of $U_{\phi}$ and $U_t$) and then perform a
final substitution for $U_{\phi}$ and $U_t$ by $(U_t)^{\pm}$.  Simplify the
resulting expression, which has the desired functional form:  $F_E(r,U^r;F_L)$.

The critical point must satisfy the conditions 
\begin{eqnarray}
\nonumber
\partial_r F_E(r_{crit},U^r_{crit};F_L) &=& 0,\\
\nonumber
\partial_{U^r} F_E(r_{crit},U^r_{crit};F_L)&=& 0,
\end{eqnarray}
and the conserved quantity $F_E$ must also satisfy the constraint at
the outer boundary
\begin{eqnarray}
\nonumber
F_E(r_{crit},U^r_{crit};F_L) &=& F_E(r_{mso},0;F_L).
\end{eqnarray}
These three conditions allow us to solve for $r_{crit}$, $U^r_{crit}$, and 
$F_L$. 

\begin{figure}[hb]
     \epsscale{0.4}
     \plotone{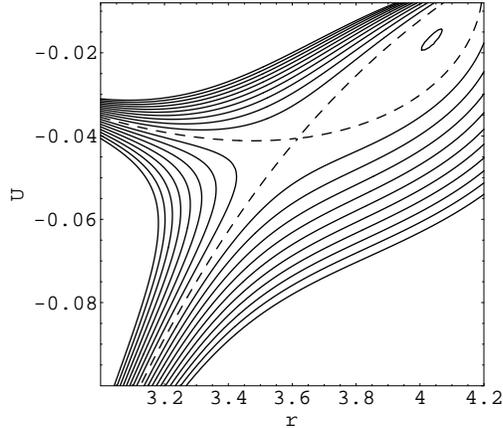}
     \caption{\label{FEsurf} 
     Contour plot of the function $F_E(r,U^r;F_L)$. The saddle point is
     located at the intersection of the dashed contours. The ingoing
     solution follows the steeper of the dashed contours; the outgoing
     solution follows the other dashed contour.} 
\end{figure}

In order to find the critical point, we will eventually use
the FindRoot function of Mathematica, but we first need to obtain a rough
location for the saddle point.  To do this, we produce a coarse contour plot
over the range $r_{horizon} < r < r_{mso}$ and a large range of $U^r$, say $-0.5
< U^r < 0$ for a given choice of root of $(U_t)^{\pm}$.  If the contour plot
does not show evidence of a saddle point, we choose the other root.  Figure
\ref{FEsurf} illustrates this for our specific test problem.  The saddle point
is shown to lie near $(r_{crit},U^r_{crit}) \approx (3.65,-0.04)$. This
is then used as an initial guess for the Mathematica FindRoot function, from
which the parameters used in the test were obtained.

To generate an initial state within our code, we apply a numerical root-finder
to the energy function $F_E(r,U^r;F_L)$ (coded using the Mathematica FortranForm
command), and generate a numerical solution for $U^r$ (using the parameters in
the text).  Once $U^r$ is found, we construct $U_t$ and $U_\phi$
using the algebraic expressions obtained from Mathematica (also coded using
FortranForm).  Once these variables have been initialized the remainder of the
code variables can be obtained in a straightforward manner.


\end{document}